\documentclass[12pt]{iopart}
\expandafter\let\csname equation*\endcsname\relax
\expandafter\let\csname endequation*\endcsname\relax
\usepackage{amsmath,amssymb,bm,dsfont}
\usepackage{cite,txfonts,comment}
\usepackage{graphicx,color}
\usepackage{braket}
\usepackage[colorlinks=true,linkcolor=blue,urlcolor=blue,citecolor=blue]{hyperref}

\usepackage{etoolbox}
\makeatletter
\def\@mkboth#1#2{}
\newlength\appendixwidth
\preto\appendix{\addtocontents{toc}{\protect\patchl@section}}
\newcommand{\patchl@section}{%
  \settowidth{\appendixwidth}{\textbf{Appendix }}%
  \addtolength{\appendixwidth}{1.5em}%
  \patchcmd{\l@section}{1.5em}{\appendixwidth}{}{\ddt}%
}
\makeatother

\begin{document}

\title{Critical non-Hermitian topology induced quantum sensing}

\author{S~Sarkar$^{1,2}$, F~Ciccarello$^{3,4}$, A~Carollo$^{3}$ and A~Bayat$^{1,2}$}

\address{${}^1$ Institute of Fundamental and Frontier Sciences, University of Electronic Science and Technology of China, Chengdu 611731, China}
\address{${}^2$ Key Laboratory of Quantum Physics and Photonic Quantum Information, Ministry of Education, University of Electronic Science and Technology of China, Chengdu 611731 , China}
\address{${}^3$ Universit`a degli Studi di Palermo, Dipartimento di Fisica e Chimica – Emilio Segr`e, via Archirafi 36, I-90123 Palermo, Italy}
\address{${}^4$ NEST, Istituto Nanoscienze-CNR, Piazza S. Silvestro 12, 56127 Pisa, Italy}

\ead{saubhik.sarkar@uestc.edu.cn}
\ead{francesco.ciccarello@unipa.it}
\ead{angelo.carollo@unipa.it}
\ead{abolfazl.bayat@uestc.edu.cn}

\begin{abstract}
Non-Hermitian physics predicts open quantum system dynamics with unique topological features such as exceptional points and the non-Hermitian skin effect. 
We show that this new paradigm of topological systems can serve as probes for bulk Hamiltonian parameters with quantum-enhanced sensitivity reaching Heisenberg scaling.
Such enhancement occurs close to a spectral topological phase transition, where the entire spectrum undergoes a delocalization transition.
We provide an explanation for this enhanced sensitivity based on the closing of point gap, which is a genuinely non-Hermitian energy gap with no Hermitian counterpart.
This establishes a direct connection between energy-gap closing and quantum enhancement in the non-Hermitian realm.
Our findings are demonstrated through several paradigmatic non-Hermitian topological models in various dimensions and potential experimental implementations.
\end{abstract}

\maketitle

\section{Introduction}

\begin{figure}[b]
\centering
\includegraphics[width = 0.7\linewidth]{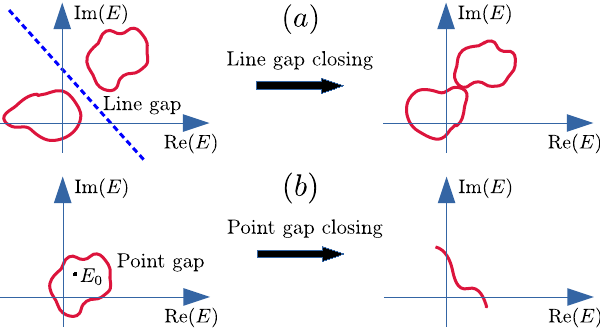}
\caption{Energy gaps of NH Hamiltonians in complex energy plane. (a) Line gap (blue dashed line) separating two NH bands (red loops). Line gap closes when the loops merge. (b) Point gap inside a spectral loop (e.g.~$E_0$). Point gap closes when the loop shrinks to an arc.}
\label{Schematic_gap}
\end{figure}

Non-Hermitian (NH) Hamiltonians are a longstanding tool for describing open system dynamics. 
Nonetheless, only in recent years it was realized that NH systems can exhibit fundamentally new phenomena in both classical and quantum systems with topological nature~\cite{Ashida2020Non}.
These include the occurrence of exceptional points, the NH skin effect and violation of bulk-boundary correspondence~\cite{Bergholtz2021Exceptional, Okuma2023Non}. 
Such effects stem from the unique topology caused by energy gaps in the complex spectrum of NH systems, which can be of two different types: the line gap and the point gap~\cite{Kawabata2019Symmetry}.
A line gap is a reference line in the complex energy plane that separates different bands, see figure~\ref{Schematic_gap}(a). 
On the other hand, a point gap is a reference energy that is not reachable by any eigenstates of the NH Hamiltonian, see figure~\ref{Schematic_gap}(b).
When the spectrum forms a loop, any interior point is a point gap with non-trivial spectral topology.
In this case, the topological invariant is the nonzero spectral winding number~\cite{Kawabata2019Symmetry}.

A major feature of NH physics is the extreme sensitivity of the spectrum to boundary conditions, e.g.~the coupling at the boundary that can smoothly change from Open Boundary Conditions (OBC) to Periodic Boundary Conditions (PBC). 
This phenomenon has been exploited for quantum-enhanced sensing protocols with NH probes to estimate the coupling at the boundary~\cite{Wiersig2014Enhancing, Liu2016Metrology, Langbein2018No, Lau2018Fundamental, Zhang2019Quantum, Chen2019Sensitivity, budich2020non, Koch2022Quantum, Schomerus2020Nonreciprocal, McDonald2020Exponentially, Hodaei2017Enhanced, Yu2020Experimental, Wang2020Petermann,Edvardsson2022Sensitivity, Ding2023Fundamental}.
In contrast, the {\it bulk} Hamiltonian parameter estimation problem is hardly explored.
Most topological features are associated with the boundary and the conventional bulk-boundary correspondence breaks down in the presence of NH skin effect.
The capability of such systems for detecting bulk parameters is therefore nontrivial and can provide new insight into the origin of quantum-enhanced sensitivity.
Devising such quantum sensors based on NH topology is highly desirable as this would advance quantum sensing in the presence of decoherence.

In general, a sensing protocol aims to enhance the accuracy in the estimation of an unknown parameter $\lambda$, which is encoded in the state of a (classical or quantum) system called {\it probe}. 
Accuracy is quantified by the statistical standard deviation $\delta\lambda$, which is lower-bounded according to the well-known Cram\'er-Rao inequality, $\delta \lambda {\ge} 1/\sqrt{\mathcal{M} F}$.
Here, $\mathcal{M}$ is the number of independent trials and $F$ is the Fisher information~\cite{paris2009quantum}, which in general grows with the probe size $L$ as $F {\sim} L^b$ with a positive exponent $b$.
The maximal scaling achievable with a classical probe is $b {=} 1$ (shot noise limit or standard limit).
Using a quantum probe, instead, can result in $b {>} 1$. 
Such quantum-enhanced sensitivity typically exploits genuine quantum features such as superposition and entanglement. 
For instance, using Greenberger-Horne-Zeilinger entangled states one can achieve $b {=} 2$ (Heisenberg limit)~\cite{giovannetti2004quantum, giovannetti2006quantum, frowis2011stable, PhysRevA.97.042112, kwon2019nonclassicality}.
However, these probes may suffer from extreme detrimental effects in the presence of decoherence~\cite{demkowicz2012elusive} or perturbations in the system~\cite{Pasquale2013Quantum}.
These problems can be tackled using quantum many-body sensors, while still achieving enhanced sensitivity near quantum criticality.
This was shown for various types of quantum phase transitions such as second-order ~\cite{zanardi2006ground, zanardi2007mixed, zanardi2008quantum, invernizzi2008optimal, gu2010fidelity, gammelmark2011phase, skotiniotis2015quantum, rams2018limits, chu2021dynamic, liu2021experimental, montenegro2021global}, Floquet~\cite{mishra2021driving, Mishra2022Integrable}, Stark-localization~\cite{he2023stark,yousefjani2023long}, boundary time crystal~\cite{montenegro2023quantumenhanced,iemini2023floquet}, and topological~\cite{Sarkar2022Free} phase transitions.
All of these transitions bear one common feature, which is the closing of an energy gap. 
Accordingly, it is natural to conjecture that {\it energy-gap closing} is the fundamental ingredient for quantum enhancement. 
Therefore, it is conceptually important to assess the possibility of enhanced sensitivity in NH systems due to their unique energy gaps (see figure~\ref{Schematic_gap}) that have no correspondence in Hermitian systems.
 
In this work, we show that NH topological systems can indeed achieve quantum-enhanced sensitivity with Heisenberg scaling as a direct consequence of point gap closing (see figure~\ref{Schematic_gap}(b)). 
Our results are established through the study of several prototypical NH topological systems in various dimensions and even with disorder. 
We first provide the necessary overview of single parameter estimation and the spectral topology in NH systems in sections~\ref{sensing} and~\ref{NHSE}, respectively.
To show the unique effect of NH topology, we report the enhanced sensitivity in the single-band Hatano-Nelson model in section~\ref{HN}.
We further move towards multi-band systems by considering a NH extension of the Su–Schrieffer–Heeger (SSH) model in section~\ref{SSH}.
To show the robustness of sensing against disorder, we look at the Aubrey-Andr\'{e}-Harper (AAH) model in section~\ref{AAH}.
Section~\ref{2D} reports the sensitivity performance in a 2D NH Chern insulator model.
Section~\ref{basis} confirms that the optimal measurement to achieve the ultimate sensitivity can be done in the experimentally relevant position basis.
In section~\ref{experiment}, we point out how the enhancement reported in this paper can be observed in ongoing experiments.
We report the enhanced sensitivity in NH systems with line gaps in section~\ref{line gap} before we conclude in section~\ref{conclusion}.

\section{Parameter estimation} 
\label{sensing}

To estimate a single unknown parameter $\lambda$, this is first encoded in a probe's quantum state $\rho_\lambda$. 
Next, a measurement on the system is performed, the outcomes of which are processed by a statistical estimation algorithm. 
The measurement is described by a set of projective operators $\{\Pi_n\}$ with the probability of the $n$th outcome given by $p_n(\lambda) {=} \text{Tr}\left[\rho_\lambda \Pi_n\right]$. 
With this classical probability distribution, the lower bound of accuracy is associated with the basis-dependent classical Fisher information, $F_C {=} \sum_n p_n (\partial_\lambda\log p_n)^2$~\cite{paris2009quantum}. 
The maximum of $F_C$ over all possible measurements is called the quantum Fisher information (QFI), which achieves the ultimate precision bound. 
QFI can be expressed as $F_Q {=} \text{Tr}\left[\mathcal{L}_{\lambda}^2 \rho_\lambda \right]$, with the symmetric logarithmic derivative operator $\mathcal{L}_\lambda$ defined implicitly as $\partial_{\lambda}\rho_{\lambda} {=} (\rho_\lambda \mathcal{L}_\lambda {+} \mathcal{L}_\lambda \rho_\lambda)/2$. 
For pure states  $\rho_{\lambda} {=} \ket{\psi_{\lambda}} \bra{\psi_{\lambda}}$ one gets $\mathcal{L}_{\lambda} {=} 2 \partial_{\lambda} \rho_{\lambda}$, and hence $F_Q {=} 4\left(\braket{\partial_\lambda \psi_{\lambda}|\partial_\lambda \psi_{\lambda}} - |\braket{\partial_\lambda \psi_{\lambda}|\psi_{\lambda}}|^2 \right)$~\cite{paris2009quantum}. 
While the optimal measurement basis to obtain the ultimate precision bound is not unique, one choice is always given by the $\mathcal{L}_\lambda$ eigenbasis. 

While the formalism laid out above has been developed for Hermitian systems, one can still use it in non-Hermitian cases under special circumstances that have been considered in this work.
The probe state for us is a particular right eigenstate of a NH Hamiltonian, which is normalized by conventional norm so that measurement procedures produce normalized probability distributions.
This is a standard practice with pure states in NH domain, both theoretically~\cite{Alipor2014Quantum, Yu2023Quantum} and experimentally~\cite{Xiao2020Non, Yu2024Toward}.
This enables us to define a valid density operator upon which the standard procedure of defining QFI~\cite{paris2009quantum, braunstein1994statistical} can be carried out, resulting in the same expression mentioned above.

\section{Spectral topology and NH skin effect}
\label{NHSE}

In Hermitian systems, topological phase transitions are accompanied by a band gap closing under PBC and correspond to the emergence of gapless edge states under OBC~\cite{Hasan2010Colloquium}. 
The Hermitian band gap closing is topologically equivalent to the line gap closing in NH systems~\cite{Kawabata2019Symmetry} [see figure~\ref{Schematic_gap}(a)).
This suggests that quantum-enhanced sensitivity can also be achieved near a NH line gap closure. 
However, NH systems also support a unique topology of the complex spectra, which has no correspondence in Hermitian systems. 
In 1D lattice systems with PBC, the spectra can form loops due to the cyclic nature of the quasi-momentum. 
Point gaps in the interior of these loops (see figure~\ref{Schematic_gap}(b)) are associated to a non-zero spectral winding number~\cite{Kawabata2019Symmetry}.
Presence of a spectral loop corresponds to NH skin effect, i.e.~edge-localization of all bulk eigenstates under OBC~\cite{Yao2018Edge, Martinez2018Non}.
When the Hamiltonian parameters are varied in a way that the PBC loops contract to arcs, the NH skin effect vanishes~\cite{Borgnia2018Non, Okuma2020Topological, Zhang2020Correspondence}. 
We refer to this as \emph{point gap closing}, where the spectral winding number becomes zero and signals a \emph{spectral topological phase transition}.
In higher dimensions, the eigen-energies can form spectral areas as the quasi-momenta cover the Brillouin zone (e.g.~see the blue regions in Fig~\ref{Fig_QWZ}(a)). 
Finite spectral area and arcs with zero area, respectively, correspond to presence and absence of NH skin effect for systems with OBC~\cite{Zhang2022Universal}, although direct connection with point gap topology is yet to be established~\cite{Okuma2023Non}. 
However, the spectral area can be thought to be constituted of a continuum of loops, each of which are formed by varying the momentum in only one spatial direction while keeping the other momenta fixed. 
Each of these loops can be associated with a spectral winding number and point gap~\cite{zhong2021nontrivial}. 
By tuning the Hamiltonian parameters, it is possible to contract all the loops into arcs at once, which we refer to as \emph{simultaneous point gap closing}.
As a result, the total spectral area can collapse into an arc (e.g.~see the red lines in Fig~\ref{Fig_QWZ}(a)), signaling a vanishing NH skin effect for any OBC geometry~\cite{Zhang2022Universal}. 
The other possibility, where the spectral area is nonzero but constitutes of arcs, corresponds to vanishing NH skin effect for particular OBC geometries.
See~\ref{Appendix_2D} for more details.
In all such cases, changes in the nature of the OBC eigenstates can be used for sensing the parameter.

In this work, we consider first order skin effect, where almost all the eigenstates are edge-localized.
When the point gap closes, all these skin states go through a sharp change and become delocalized, which makes them potential candidates for quantum-enhanced sensitivity.
However, from a practical perspective, the most relevant state for evolution with a NH Hamiltonian is the right eigenstate with OBC that has the largest imaginary part in its energy as this will be the only state surviving in the long time limit~\cite{Gong2018Topological, Lieu2019Non, Panda2020Entanglement, Banerjee2022Chiral}. 
We take this state, after normalizing, as the probe state for most cases considered in this work.
We note that, point gap closing can also occur in NH systems with higher order skin effects~\cite{Kawabata2020Higher} although the number of skin states are significantly diminished.
We focus on first order skin effect in this work as here the transition happens almost across the whole spectrum.
This is promising from an experimental view as the effect of enhancement in the single particle energy levels should have dynamical imprints.
In the following, we investigate the sensing capability of an extensive set of NH systems and show that point gap closing indeed results in quantum-enhanced sensitivity.

\section{Single-band case} 
\label{HN}

\begin{figure}[t]
\centering
\includegraphics[width=0.7\linewidth]{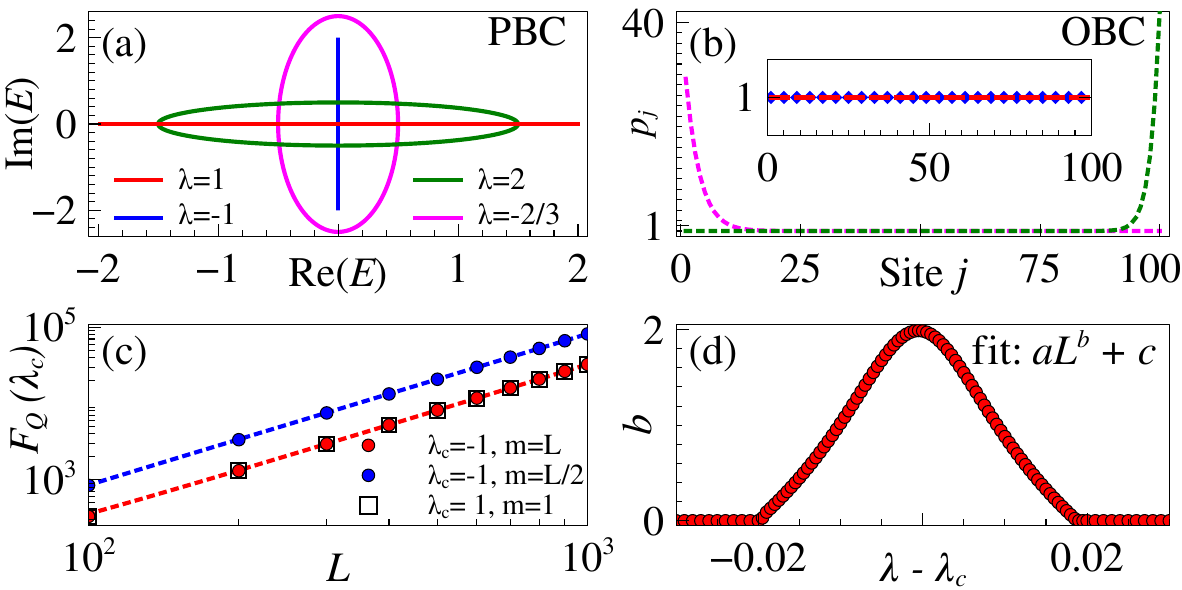} 
\caption{Hatano-Nelson model with $\lambda = J_R/J_L$ in equation \eqref{HN_ham}. (a) PBC spectrum. (b) Cumulative site-population for OBC eigenstates shows the NH skin effect and its absence (inset). (c) Quadratic scaling of QFI at the transition $\lambda {=} \lambda_c$. (d) QFI Scaling exponent near transition.}
\label{Fig_HN}
\end{figure}

We first study the Hatano-Nelson model which is a 1D chain of size $L$, with the Hamiltonian~\cite{Hatano1996Localization, Hatano1997Vortex} 
\begin{align}
    H_{\rm HN} = \sum_{j} \left( J_L \ket{j} \bra{j{+}1} + J_R \ket{j{+}1} \bra{j}\right) ,
    \label{HN_ham}
\end{align}
where $J_L$ and $J_R$ are real asymmetric hopping terms towards left and right, respectively. 
The parameter to be estimated is $\lambda {=} J_R/J_L$. 
At $\lambda{=}\lambda_c{=}{\pm} 1$ the system's point gap closes, signalled by a change in the winding number (see~\ref{Appendix_w}).
Note that while at $\lambda_c {=} 1$ the Hamiltonian becomes Hermitian, $\lambda_c {=} {-}1$ makes it anti-Hermitian. 
Under PBC, this Hamiltonian can be diagonalized in the quasi-momentum $\ket{k}$ basis, with a single-band spectrum $E_k {=} (J_L {+} J_R) \cos{k} + i (J_L {-} J_R) \sin{k}$,  with $k {\in} [0, 2\pi)$. 
In figure~\ref{Fig_HN}(a), we plot this spectrum in the complex energy plane which forms closed loops for $\lambda {\neq} \lambda_c$. 
Point gap closing occurs at the critical values $\lambda {=} \lambda_c$, for which the loops contract to a line. 
The spectrum changes dramatically if one considers OBC instead. 
Here the analytical solutions can be obtained by employing the generalized Brillouin zone (GBZ) formalism where the translational invariance in the large system size limit is exploited to write down an ansatz for a bulk eigenstate as $\ket{\psi} {=} \sum_{j} \beta^j \phi  \ket{j}$~\cite{Yao2018Edge, Yao2018Non, Yokomizo2019Non, Yokomizo2023Non}.
See~\ref{Appendix_GBZ} for more details.
For the Hatano-Nelson model, there are two solutions for $\beta$ satisfying $|\beta| {=} \sqrt{|\lambda|}$.
The energies are given by $E_m {=} 2 (J_R J_L)^{1/2} \cos{\frac{m \pi}{L+1}}$, with $m {\in} [1,L]$, and the corresponding eigenstates are $\ket{E_m} {=} \sum_{j} w_{j,m} \ket{j}$ in which (see~\ref{Appendix_NH}, \cite{Yokomizo2019Non, guo2021exact})
\begin{align}
    w_{j,m} = \mathcal{N}_m \lambda^{j/2} \sin{\frac{m \pi}{L+1}},
    \label{eq:alpha}
\end{align}
with the normalization factor $\mathcal{N}_m$. 
Due to the presence of $\lambda^{j/2}$ in equation (\ref{eq:alpha}), for $\lambda {\neq} \lambda_c$, every eigenstate is exponentially localized at either edge with localization length ${\sim} 1/\log(|\lambda|)$. 
In contrast, the eigenstates become  delocalized at $\lambda_c$ and equivalently $|\beta| {\to} 1$.
To see this more explicitly one can define the cumulative population at site $j$ as $p_j {=} \sum_m |w_{j,m}|^2$ which is displayed in figure~\ref{Fig_HN}(b), for the same $\lambda$ values as in figure~\ref{Fig_HN}(a). 
The localization of $p_j$ clearly shows that whenever the PBC spectrum forms a loop, all the corresponding OBC eigenstates localize at the edges, which is the NH skin effect. 
To evaluate how the sensitivity scales with the system size, we compute the QFI with respect to  $\lambda$ for different OBC eigenstates.
In figure~\ref{Fig_HN}(c), the QFI at the transition $F_Q(\lambda_c)$ is plotted for the three representative  eigenstates indexed by $m{=}1, 2, L$. 
The scaling at the transition $F_Q(\lambda_c) {\sim} aL^b$, with $b{\approx}2$, clearly shows critically-enhanced Heisenberg scaling for all eigenstates. 
The analytical derivation of this scaling is presented in~\ref{Appendix_NH}.
Unlike exponent $b$, coefficient $a$ depends on the choice of the eigenstate. 
To study the sensitivity across the transition, in figure~\ref{Fig_HN}(d) we plot the exponent $b$ versus $\lambda {-} \lambda_c$ with $\lambda_c {=} {-}1$ for the eigenstate  with $m{=}L$, corresponding to the largest imaginary energy. 
Other eigenstates behave similarly. 
Deviation from the transition causes the exponent $b$ to eventually vanish, signaling the emergence of a localized phase. 

\section{Two-band case} 
\label{SSH}

\begin{figure}[t]
\centering
\includegraphics[width=0.7\linewidth]{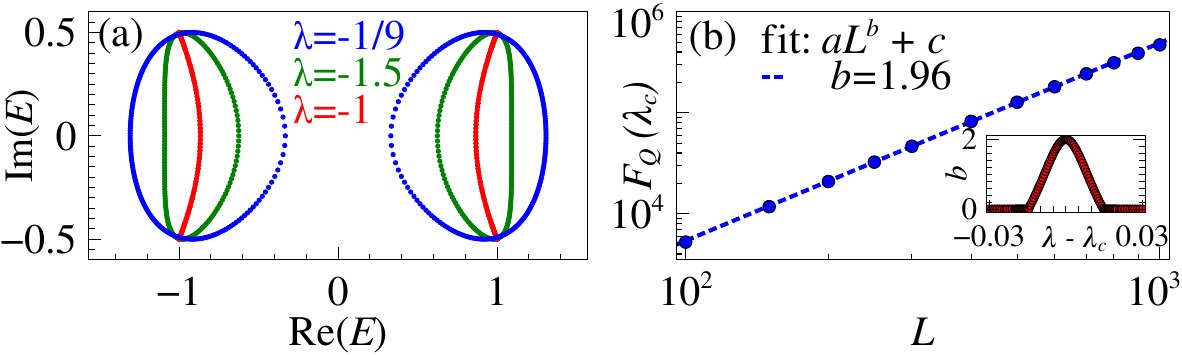} 
\caption{Non-Hermitian SSH model with $\lambda = J_{1R}/J_{1L}$ in equation \eqref{SSH_SL_ham}. Here $J_2 {=} 1$. (a) PBC spectrum. (b) Quadratic scaling of QFI at the transition. (Inset) QFI Scaling exponent near transition.}
\label{Fig_SSH}
\end{figure}

Our next focus is the NH version of the Su–Schrieffer–Heeger (SSH) model for an 1D chain with $L$ cells, each with two sites $A$ and $B$~\cite{Yao2018Edge}. 
The Hamiltonian is
\begin{align}
    H_{\rm SSH} = \sum_{j}  \Big( J_{1L} \ket{j, A} \bra{j, B} + J_{1R} \ket{j, B} \bra{j, A} + J_2 (\ket{j, A} \bra{j{+}1, B} + \ket{j{+}1, B} \bra{j, A}) \Big) ,
    \label{SSH_SL_ham}
\end{align}
with asymmetric intra-cell hoppings $J_{1L}$ and $J_{1R}$ and inter-cell hopping $J_2$. 
The parameter of interest is $\lambda{=}J_{1R}/J_{1L}$.  
Here, due to the presence of two sub-lattices ($A$ and $B$), the PBC spectrum generally forms two loops in the complex energy plane (similar to two bands in the Hermitian case).  
Consequently, this model shows a rich phase diagram as a function of the Hamiltonian parameters~\cite{Halder2023Properties}.
This includes: (i) merging of two loops into a single one, signaling line gap closing, equivalent to the standard Hermitian band gap closing; and (ii) shrinking of loops into arcs (as in the previous model), signaling point gap closing (red curve in figure~\ref{Fig_SSH}(a)).  
As before, the presence of the PBC spectral loops correspond to localized OBC bulk states, namely NH skin effect.
The GBZ ansatz here results again in two possible localization parameters satisfying $|\beta| {=} \sqrt{|\lambda|}$~\cite{guo2021exact}, see~\ref{Appendix_SSH} for more details.
 In figure~\ref{Fig_SSH}(a), we set $J_2{=}1$ and plot the PBC spectrum for different values of $\lambda$.
The spectral loops collapse into arcs when $|\lambda| {=} 1$, where the point gap closes and the OBC eigenstates delocalize as $|\beta| {\to} 1$.
To investigate the sensitivity at the transition, we calculate the QFI of the OBC eigenstate with largest imaginary energy (i.e.~the steady state) at $\lambda_c {=} {-}1$.
There are two eigenstates with largest imaginary energy near $\lambda_c$, with equal and opposite real energies.
We have numerically checked that the QFI at the transition behaves similarly for both the states as well as for their superpositions.
In figure~\ref{Fig_SSH}(b), the QFI is plotted against the system size $L$ for the state with largest imaginary energy and negative real energy.
Again one observes the critically-enhanced Heisenberg scaling $F_Q(\lambda_c) {\sim} L^b$, with $b {\approx} 2$. 
As the inset shows, the enhancement decreases away from $\lambda_c$.
Note that the enhanced sensitivity is generic to all eigenstates and the choice of this specific eigenstate is motivated by its dominance in the long-time dynamics. 

\section{Disordered case} 
\label{AAH}

\begin{figure}[t]
\centering
\includegraphics[width=0.7\linewidth]{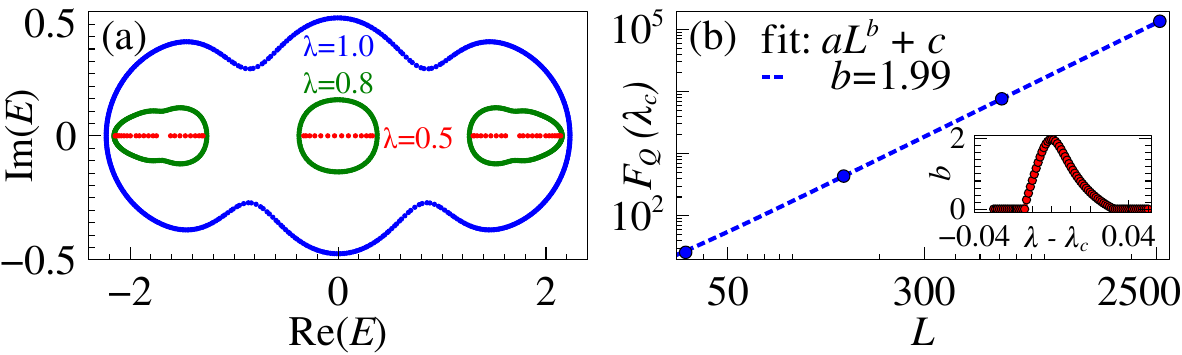} 
\caption{Non-Hermitian AAH model with $\lambda = h$ in equation \eqref{AAH_ham}. Here $J {=} V {=} 1$, $\alpha {=} (\sqrt{5}{+}1)/2$, $\theta {=} 0$. (a) PBC spectrum. (b) Quadratic scaling of QFI at the transition. (Inset) QFI Scaling exponent near transition.}
\label{Fig_AAH}
\end{figure}

We now consider a NH extension of the Aubrey-Andr\'{e}-Harper (AAH) Hamiltonian, given by~\cite{Longhi2019Topological}
\begin{align}
    H_{\rm AAH} = \sum_{j} \Big( J (\ket{j} \bra{j{+}1} + \text{h.c.}) + V \cos{(2 \pi \alpha j {+} \theta {+} i h)}\ket{j} \bra{j} \Big) \,.
    \label{AAH_ham}
\end{align}
Here $J$ is a standard hopping parameter, while $V$ is the disorder amplitude, $\alpha$ the (irrational) quasi-frequency, $\theta$ a real phase and $h$ is the strength of an imaginary phase which has to be estimated. 
The value $h_c {=} \log{(2J/V)}$ is a multi-critical point for both parity-time (PT) symmetry breaking and localization transitions, captured by a spectral winding number with $\theta$ running from $0$ to $2\pi$ (see~\ref{Appendix_w} for more details). 
In the thermodynamic limit, this number vanishes for $h {<} h_c$ while the eigenstates are delocalized with real PBC spectrum (unbroken PT phase).
For $h {>} h_c$, instead, the winding number is $-1$ and the eigenstates are now localized with complex PBC spectrum (broken PT phase) that can form spectral loops (see figure~\ref{Fig_AAH}(a)).
Due to lack of translational invariance, the GBZ formalism does not apply here.
To avoid finite-size effects, we follow the standard numerical procedure~\cite{Jiang2019Interplay, wei2019fidelity} for choosing the irrational parameter $\alpha$. 
Accordingly, for a system size $L {=} \mathcal{F}_{n}$ (with $\mathcal{F}_{n}$ as the $n$-th Fibonacci number), $\alpha$ is approximated by $\mathcal{F}_{n+1}/\mathcal{F}_{n}$.
As $n {\to} \infty$, $\alpha$ converges to the golden ratio (\!$\sqrt{5} {+} 1)/2$.
To infer the scaling of the QFI with respect to $h$ (i.e.~for $\lambda {=} h, \lambda_c {=} h_c$) versus system size, we vary $n$ between 9 (corresponding to $L {=} 34$) and 18 (corresponding to $L {=} 2584$).  
In figure~\ref{Fig_AAH}(b) we plot $F_Q(\lambda_c)$ as a function of $L$, which again shows Heisenberg scaling $F_Q(\lambda_c) {\sim} L^b$, with $b {\approx} 2$. 
This scaling is in agreement with that of the fidelity susceptibility in the Hermitian case~\cite{wei2019fidelity}. 
In the inset we show the behavior of $b$ away from the transition.
The above results also show the robustness of sensors based on topological systems against local perturbations, as the Hatano-Nelson model with random disorder can be mapped to the Hamiltonian $H_{\rm AAH}$~\cite{Longhi2019Topological}.

\section{Two-dimensional case} 
\label{2D}

\begin{figure}[t]
\centering
\includegraphics[width=0.7\linewidth]{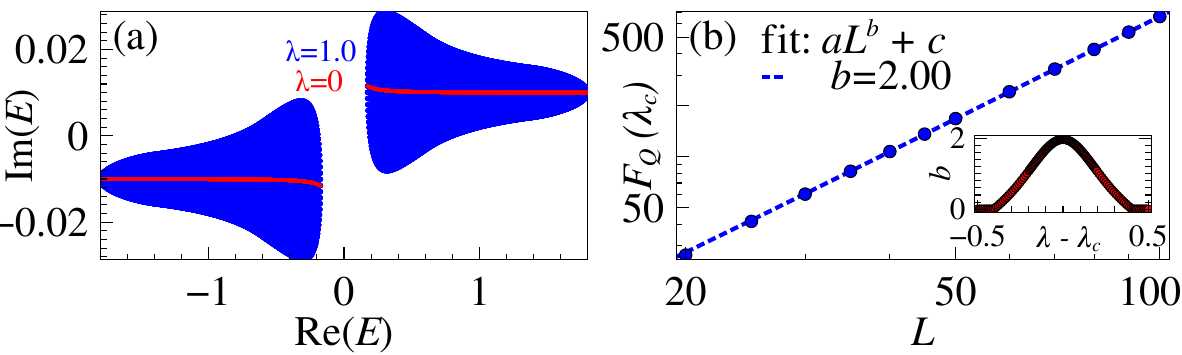} 
\caption{Non-Hermitian QWZ model with $\lambda = t_1/m_z$ in equation \eqref{Ch_ham}. (a) PBC spectrum. The fixed parameters for a $L {\times} L$ lattice are, $m_z {=} 1, t_2 {=} 0.2, \gamma_x {=} 0.1, \gamma_y {=} 0.1, \gamma_z {=} 0.01, L {=}200$. (b) Quadratic scaling of QFI at the transition for OBC along both directions. (Inset) QFI Scaling exponent near the transition for the same OBC.}
\label{Fig_QWZ}
\end{figure}

As a prototypical 2D topological model, we consider a NH version of the Qi-Wu-Zhang (QWZ) Chern insulator~\cite{Shen2018Topological, Yao2018Non, qi2006topological}. 
This is a bipartite square lattice with Bloch Hamiltonian at momentum $\boldsymbol{k}=(k_x,k_y)$ given by 
\begin{align}
    H_{\rm QWZ} (\boldsymbol{k}) = (2 t_1 \sin{k_x} + i \gamma_x) \sigma_x + (2 t_1 \sin{k_y} + i \gamma_y) \sigma_y + [m_z - 2 t_2 (\cos{k_x} + \cos{k_y}) + i \gamma_z] \sigma_z,
    \label{Ch_ham}
\end{align} 
where  $\sigma_{x,y,z}$ are the Pauli matrices, $t_1, t_2$ are hopping parameters, $m_z$ is an onsite term (which is fixed to unity), and $\gamma_{x,y,z}$ are the imaginary terms making the system NH. 
We note that, since there is no unique definition of spectral winding number in 2D systems, finding the value of the critical parameter for the transition requires some care.
We now briefly outline how to determine the parameters driving the transition (see~\ref{Appendix_2D} for further details).
The PBC spectrum generically consists of two spectral areas corresponding to two bands, shown as blue areas in figure~\ref{Fig_QWZ}(a). 
Each band in turn can be decomposed into spectral loops each forming when $k_y$ (for example) varies in the first Brillouin zone, while $k_x$ is held fixed.
Accordingly, a $k_x$-dependent spectral winding number~\cite{zhong2021nontrivial} can be written for each loop relying on their point gap topology. The parameter values for the transition are then found using the same GBZ formalism as in 1D models.
The simultaneous point gap closing occurs at three independent points, $t_1/m_z {=} 0, t_2/m_z {=} 0$, and $\gamma_y/m_z {=} 0$ (the other Hamiltonian parameters being fixed to arbitrary values).
Further numerical inspection reveals that only at $t_1/m_z {=} 0$, the contractions happen in such a way that they collectively also form an overall arc for the PBC spectrum (see red curve in figure~\ref{Fig_QWZ}(a)). 
Hence, the collapse of the spectral area corresponds to vanishing NH skin effect for any OBC geometry~\cite{Zhang2022Universal}.
Choosing $\lambda {=} t_1/m_z$ with $\lambda_c {=} 0$ and OBC along both directions, we calculate the QFI for the OBC eigenstate with largest imaginary energy (i.e.~the steady state).
For an $L {\times} L$ lattice, we find the scaling of QFI at transition as $F_Q(\lambda_c) {\sim} L^b$, with $b {\approx} 2$.
The inset shows how $b$ decreases away from $\lambda_c$.
The other simultaneous point gap closing instances correspond to vanishing of skin effect if OBC is only along $y$, and similar scaling of QFI is found for that OBC geometry.

\section{Optimal measurement basis} 
\label{basis}

These prototypical examples regarding the NH spectral topology establish the connection between point gap closing and quantum-enhanced sensing with the physically relevant states. 
It turns out that the optimal measurement basis is given by the position basis (except for the disordered case where ${\sim} 95\%$ of QFI is obtained), which is typical for QFI near a localization transition.
The classical Fisher information in position basis closely follows the QFI, as shown in~\ref{Appendix_CFI}.
This measurement is local in nature, hence easily implementable in practice.

\section{Experimental realization}
\label{experiment}

\begin{figure}[t]
\centering
\includegraphics[width=0.7\linewidth]{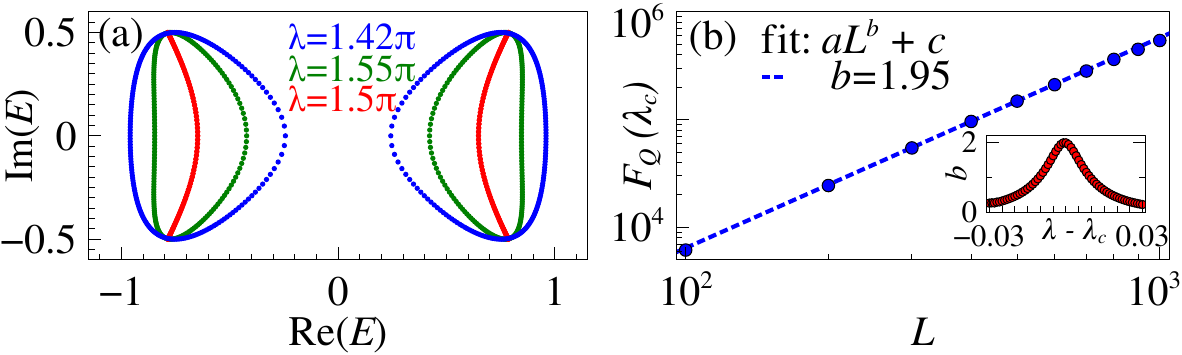} 
\caption{Non-unitary quantum walk model with $\lambda = \theta_2$ in equation \eqref{QW_ham}. Here $\theta_1 {=} \pi/4$, $\gamma {=} 0.5$ (a) PBC spectrum. (b) Quadratic scaling of QFI at the transition. (Inset) QFI scaling exponent near transition.}
\label{Fig_QW}
\end{figure}

NH models with skin effects are routinely realized in various experimental platforms such as, electric circuits~\cite{Helbig2020Generalized, Hofmann2020Reciprocal, Liu2021Non}, acoustic~\cite{Zhang2021Acoustic, Gao2022Anomalous} and photonic lattices~\cite{Weidemann2020Topological, Song2020Two}, and mechanical metamaterials~\cite{Brandenbourger2019Non, Ghatak2020Observation}. 
In the context of quantum simulations, the NH version of SSH model has been realized with lossy optical lattices~\cite{Lapp2019Engineering, Gou2020Tunable}. 
An excellent candidate for probing the point gap closing is the photonic discrete-time quantum walk experiment where the NH skin effect has been observed~\cite{Xiao2020Non, Xiao2021Observation, Lin2022Observation, Lin2022ObservationAAH}.
Specifically, in the recently conducted experiments~\cite{Xiao2020Non, Xiao2021Observation}, the internal dimensions of the quantum walker were encoded in the photon polarizations ${\uparrow, \downarrow}$ and the dynamics along an 1D lattice were governed by the non-Unitary evolution operator
\begin{align}
    U_{\rm QW} = R_{\frac{\theta_1}{2}} S_{\uparrow} R_{\frac{\theta_2}{2}} M R_{\frac{\theta_2}{2}} S_{\downarrow} R_{\frac{\theta_1}{2}} ,
    \label{QW_ham}
\end{align}
where $R_{\theta} {=} \sum_j \ket{j} \bra{j} \otimes e^{-i \theta \sigma_y}$ acts on the internal space and $S_{\uparrow (\downarrow)}$ shifts the walker to the left (right) by one lattice site if the polarization is $\uparrow (\downarrow)$. 
Non-Unitarity is implemented by the gain-loss operator $M {=} \sum_j \ket{j} \bra{j}\otimes e^{\gamma \sigma_z}$. 
The unitary coin operator $R_{\theta}$ can rotate the spin in a position-dependent way and can be implemented with different settings of half-wave plates. 
The unitary spin-dependent shift operator $S$ translates the particle by one site either way in the one-dimensional lattice and is realized with beam displacers. 
For a time-multiplexed scheme, the shift operator is realized by splitting the different polarized components with polarizing beam splitters and transferring them through two different optical fibers of different lengths. 
Different arrival times at the detectors correspond to their different shifts. 
The non-unitary spin-dependent loss operator $M$ is implemented with partially polarizing beam splitters. 
To simulate balanced gain and loss, an operator with more loss and less loss terms is used with a constant scaling factor. 
The measurement in position basis is done with avalanche photo-diodes which also resolve the spin.
$U_{\rm QW}$ gives a stroboscopic description of the dynamics under an effective NH Hamiltonian $H_{\rm eff}$, i.e.~$U_{\rm QW} {=} e^{-i H_{\rm eff}}$~\cite{Mochizuki2016Explicit}. 
The location of the point gap closing for this $H_{\rm eff}$ can be determined by the GBZ formalism.
The localization parameter $\beta$ can take two values which satisfy $|\beta| {=} \sqrt{|(\cosh{\gamma} \cos{\theta_2} {-} \sinh{\gamma})/(\cosh{\gamma} \cos{\theta_2} {+} \sinh{\gamma})|}$~\cite{Xiao2021Observation}.
As $\gamma$ is non-zero, the transition must occur at $\cos{\theta_2} {=} 0$ where $|\beta|$ is 1 and makes NH skin effect vanish for OBC.
We choose $\lambda {=} \theta_2$ with $\lambda_c {=} 3\pi/2$, and confirm that the PBC spectrum loops indeed collapse to an arc at $\lambda_c$ (see figure~\ref{Fig_QW}(a)).
The other parameters are fixed at $\theta_1 {=} \pi/4$ and $\gamma {=} 0.5$.
Figure~\ref{Fig_QW}(b) shows the algebraic scaling of QFI for the steady state with OBC at the transition again with exponent $b {\approx} 2$, while the inset shows the exponent receding to 0 as one moves away from $\lambda_c$.
Therefore, by tuning the experimental parameters close to the transition, it should be possible to probe the findings reported in this work.
Moreover, the measurements in these experiments are carried out in the position basis which is the optimal basis to achieve ultimate precision in this system.

\section{Sensing near line gap closing}
\label{line gap}

\begin{figure}[t]
\centering
\includegraphics[width=0.7\linewidth]{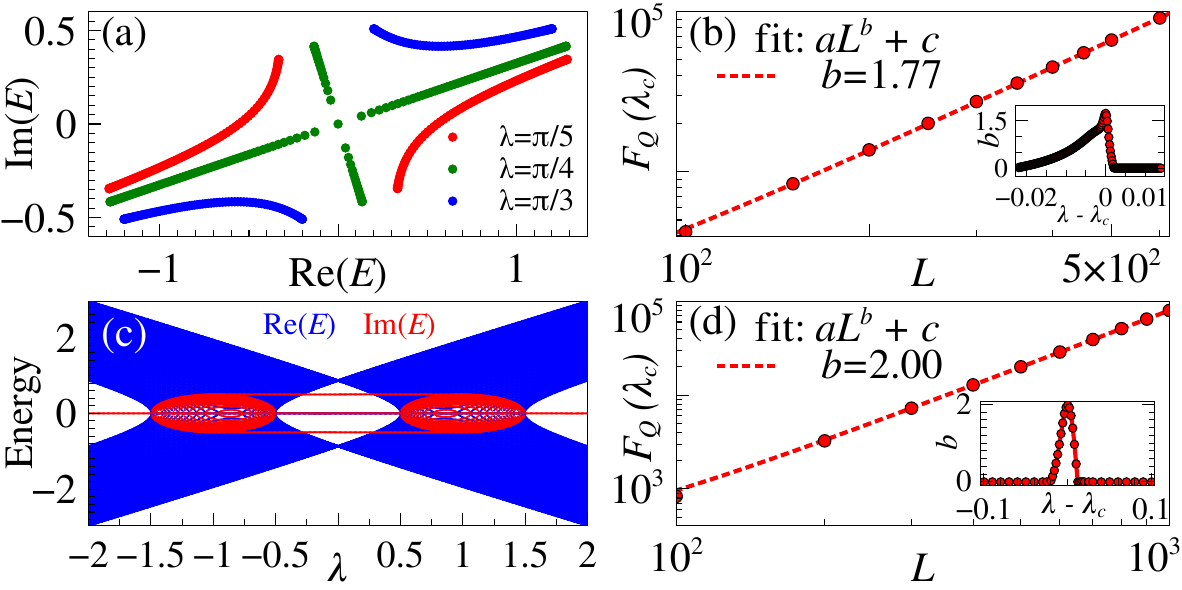} 
\caption{NH extensions of SSH model without point gap. (a) PBC spectrum of chiral inverse symmetric SSH model with $\lambda_c = \pi/4$ in Eq.~\eqref{SSH_Ch_ham}. (b) Scaling of QFI of the edge state at the transition. (Inset) QFI scaling exponent near transition. (c) OBC spectrum of PT symmetric SSH model with $\lambda = J_1/J_2$ in Eq.~\eqref{SSH_PT_ham}. Here $\delta {=} 0.5$. (d) Quadratic scaling of QFI of the edge state at the transition. (Inset) QFI scaling exponent near transition.}
\label{Fig_Line}
\end{figure}

Having established the connection between point gap closing and enhanced sensitivity, we now report on NH systems with line gap topology.
As line gap closures have direct correspondence with band gap closing Hermitian topology~\cite{Gong2018Topological, Kawabata2019Symmetry}, one expects enhanced sensitivity for the chiral edge states that emerge in the topologically non-trivial phase.
However, we find that this is not the case when the system has a non-trivial point gap.
For the SSH model in Eq.~\eqref{SSH_SL_ham}, the edge states have exactly zero energy, which makes them theoretically trackable. 
The line gap closes at  $J_1 = \pm (J_2 \pm \gamma) $ for PBC and the edge states emerge at $J_1 = \pm \sqrt{J_2^2 \pm \gamma^2}$.
This discrepancy is due to NH skin effect which causes modification of the conventional bulk-boundary correspondence~\cite{Yao2018Edge}.  
Both these cases are away from the point gap closing point and no enhanced sensing occurs.
Similar observation is found for the stripe geometry for the 2D Chern insulator in Eq.~\eqref{Ch_ham}, where the edge states actually can be chosen to have the largest imaginary energy.
This can be explained through the fact that, due to the presence of point gap, the edge state transit into bulk state, which are also localized due to NH skin effect.
We therefore look into two NH extensions of the SSH model that do not have point gap~\cite{Lieu2018Topological}.
The chiral and inversion symmetric SSH model is given by the Hamiltonian
\begin{align}
    H_{\rm SSH-I} = \sum_{j} \Big( \tilde{J}_1 (\ket{j, A} \bra{j, B} + \ket{j, B} \bra{j, A} ) 
                  + \tilde{J}_2 (\ket{j, A} \bra{j+1, B} + \ket{j+1, B} \bra{j, A}) \Big),
    \label{SSH_Ch_ham}
\end{align}
where the complex hopping parameters $\tilde{J}_1, \tilde{J}_2$ are parameterized by $\lambda$: $\tilde{J}_1 = e^{i \pi/5} \sin{\lambda}$ and $\tilde{J}_2 = \cos{\lambda}$.
Here the line gap closes at $\lambda = \pi/4$ (Fig.~\ref{Fig_Line}(a)), and the QFI for the zero-energy edge states show enhanced sensing capability (Fig.~\ref{Fig_Line}(b)).

On the other hand, the PT symmetric SSH Hamiltonian is
\begin{align}
    H_{\rm SSH-II} = \sum_{j} \Big( (J_1 \ket{j, A} \bra{j, B} + J_2 (\ket{j, A} \bra{j+1, B} + \text{H.c.}) 
                   + i \delta (\ket{j, A} \bra{j, A} - \ket{j, B} \bra{j, B}) \Big),
    \label{SSH_PT_ham}
\end{align}
with real hopping parameters and imaginary staggered potential strength $\delta$. 
Here the line gap closing occurs at $J_1 = J_2 \pm \delta$, where the PT symmetry breaking transition happens but the chiral edge states exist only for $|J_1| < |J_2|$, with zero real energy and $\pm \delta$ as the imaginary part (Fig.~\ref{Fig_Line}(c)). 
The state with energy $i \delta$ has the largest imaginary energy in the spectrum and when used to compute QFI, shows quadratic scaling at the transition $\lambda = J_1/J_2 = 1$ (Fig.~\ref{Fig_Line}(d)).

\section{Conclusion}
\label{conclusion}

In this work, we showed that quantum-enhanced sensitivity for the bulk Hamiltonian parameters of NH topological systems can be achieved with Heisenberg scaling.
This is a significant result given that these systems have so far been thought to be applicable for only sensing boundary perturbations.
Through investigation of several paradigmatic models, we found that the origin of such enhanced precision is directly connected with the closing of point gaps which affects the entire spectrum.
Remarkably, this kind of novel gap closing is a genuinely NH phenomenon with no analogue in Hermitian physics. 
We point out that realizing NH Hamiltonians require {\it open} quantum systems. 
In this respect, our work shows that, despite being usually detrimental to sensing, decoherence can be a resource for quantum-enhanced sensitivity, thus offering practical advantages.  
Finally, we showed potential implementations of our protocol in various physical platforms and point out how the enhancement can be observed in a recently conducted experiment.

\ack{F.~C.~and A.~C.~acknowledge support from University of Palermo through project “Bando Eurostart 2022". 
A.~B.~acknowledges support from the National Key R\&D Program of China (Grant No.~2018YFA0306703), the National Natural Science Foundation of China (Grants No.~12050410253, No.~92065115, and No.~12274059), and the Ministry of Science and Technology of China (Grant No.~QNJ2021167001L).}

\appendix

\section{Spectral winding number}
\label{Appendix_w}

Non-Hermitian Hamiltonians have complex eigen-energies which gives rise to an unique spectral topology that has no correspondence in Hermitian systems. The topological invariant is determined by the winding of the phase of the complex energies. This is given by the spectral winding number. In 1D lattices with translational invariance, it is defined with respect to a reference energy $E_0$ as~\cite{Kawabata2019Symmetry} 
\begin{align}
    w_{1D} (E_0) =& \frac{1}{2 \pi i} \int_{0}^{2 \pi} dk \ \partial_k \ln{(E_k - E_0)} .
    \label{w1D}
\end{align}
Spectral winding number in a 2D system for winding along $y$-direction for a given value of momentum along $x$-direction with respect to a reference energy $E_0$ is defined as~\cite{zhong2021nontrivial}
\begin{align}
    w_{2D} (k_x, E_0) =& \frac{1}{2 \pi i} \int_{0}^{2 \pi} dk_y \ \partial_{k_y} \ln{(E_{\boldsymbol{k}} - E_0)} .
    \label{w2D}
\end{align}

For the (quasi) disordered  model in equation \eqref{AAH_ham}, the winding number is defined by circulating the real phase $\theta$ of the disorder potential form 0 to $2\pi$ while taking the thermodynamic limit, i.e.~system size $L \to \infty$. Keeping the other parameter fixed, the Hamiltonian can be considered as a function of the phase, i.e.~$H_{\rm AAH} (\theta)$, and the winding number with respect to a reference energy $E_0$ is~\cite{Longhi2019Topological} 
\begin{align}
    w_{\rm AAH} (E_0) =& \frac{1}{2 \pi i} \int_{0}^{2 \pi} d\theta \ \partial_{\theta} \ln{\det(H_{\rm AAH} (\theta) - E_0)} .
    \label{wAAH}
\end{align}

\section{GBZ formalism}
\label{Appendix_GBZ}

The OBC eigenstates can be analytically studied with GBZ formalism~\cite{Yao2018Edge, Yao2018Non, Yokomizo2019Non, Yokomizo2023Non}. 
This method was developed to modify the conventional bulk-boundary correspondence, which is the central pillar of Hermitian topological physics, and breaks down in the presence of NH skin effect~\cite{Yao2018Edge, Kunst2018Biorthogonal}. 
Here, the bulk eigenstate is approximated, in the large system size limit, by the ansatz $\ket{\psi} = \sum_{j,s} \beta^j \phi_{s} \ket{j, s}$, where $j$ and $s$ denote a lattice site and internal dimension at each site, respectively.
Localization and internal state are encoded in  $\beta$  and $\phi_s$, respectively.
The eigenvalue equation results in an algebraic equation for $\beta$ and the solutions $\beta_n$ are used to write down a superposition with coefficients $c_n$ for the energy eigenstate $\ket{E} = \sum_{j,s,n} c_n \beta_n^j \phi_{n,s} \ket{j, s}$.
Imposition of boundary conditions often result in analytically tractable solutions.
In particular, the Hamiltonian parameter values at the point gap closing can be identified by looking at the limit $|\beta| {\to} 1$, which makes the eigenstates delocalize and thus NH skin effect disappears.
The GBZ formalism allows for analytical computation of the QFI in certain cases, as detailed below.
Note that $\beta^j$ in the ansatz closely resemble the phase $e^{ikj}$ appearing in Bloch's theorem. 
In the Hermitian case, $e^{ik}$ produces an unit circle in the complex plane, representing the Brillouin zone, which is generalized to a closed loop by the solutions $\beta_n$ for the NH case.
This GBZ has been successfully used to calculate the topological invariants correctly for 1D systems with tight-binding Hamiltonians, while its generalization in higher dimensions is still a topic of active research~\cite{Ashida2020Non}.

\section{QFI calculation for Hatano-Nelson model}
\label{Appendix_NH}

\begin{figure*}[t]
\centering
\includegraphics[width = 0.9\linewidth]{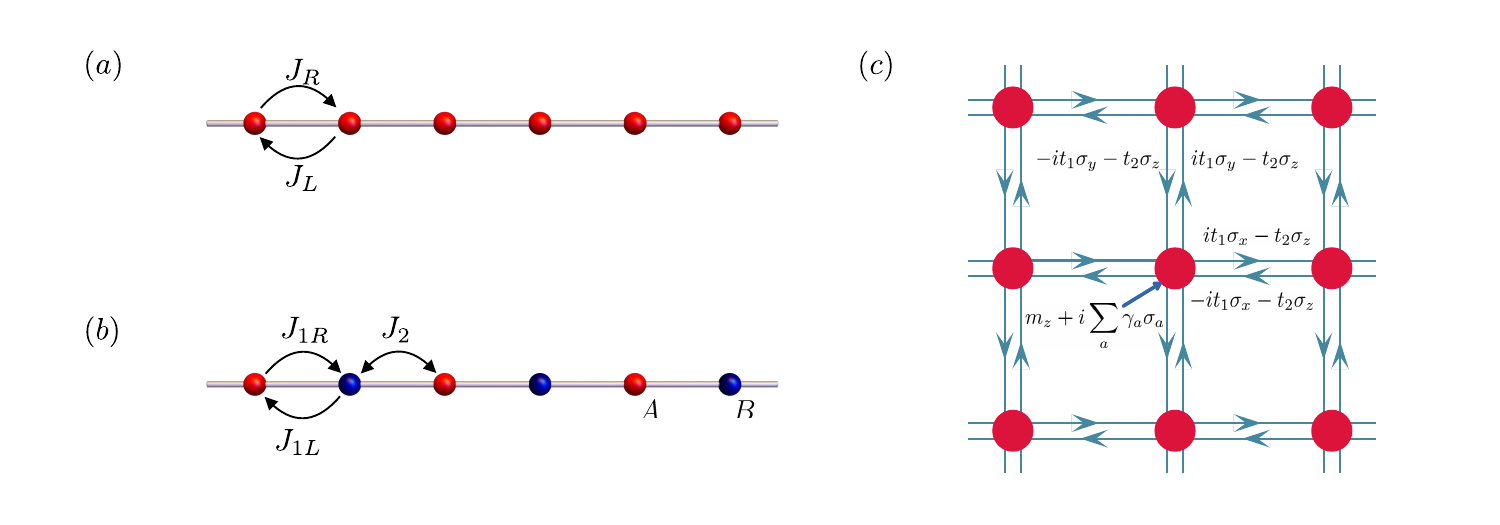}
\caption{Schematics of some of the NH models considered in this work. (a) HN model. (b) NH version of SSH model. (c) NH version of 2D Chern insulator based on QWZ model.}
\label{Schematic}
\end{figure*}

Hatano-Nelson model is a single-band model that captures the uniqueness of NH topology due to asymmetric hopping parameters to left and right on an 1D chain. The Hamiltonian is given by the equation \eqref{HN_ham}. 
The model is schematically depicted in figure~\ref{Schematic}(a).
The winding number in equation \eqref{w1D} takes values $1$ and $-1$ for $|J_R/J_L| < 1$ and $|J_R/J_L| > 1$, respectively.
The GBZ ansatz for a bulk eigenstate in the presence of NH skin effect for OBC is 
\begin{align}
    \ket{\psi} = \sum_{j} \beta^j \phi \ket{j}.
    \label{HN1}
\end{align}
The eigen-equation $H_{\rm HN} \ket{\psi} = E \ket{\psi}$ in the bulk gives
\begin{align}
    J_L \beta^2 - E \beta + J_R = 0 .
    \label{HN2}
\end{align}
The two solutions $\beta_1$ and $\beta_2$ satisfy, with definition $\lambda \equiv J_R/J_L \equiv z^2 = (r e^{i \theta})^2$,
\begin{align}
    \beta_1 \beta_2 = \frac{J_R}{J_L} = r^2 e^{i 2 \theta} \quad \text{and} \quad \beta_1 + \beta_2 = \frac{E}{J_L} .
    \label{HN3}
\end{align}
The full solution is the superposition 
\begin{align}
    \ket{\psi} = \sum_{j} (c_1 \beta_1^j \phi_1 + c_2 \beta_2^j \phi_2) \ket{j}.
    \label{HN4}
\end{align}
The eigen-equation at boundaries $(j=1,L)$ and equation \eqref{HN3} result in $\phi_1 = -\phi_2$ and
\begin{align}
    \left(\frac{\beta_1}{\beta_2}\right)^{L+1} = 1 \implies \frac{\beta_1}{\beta_2} = e^{i 2\theta_m} ,
    \label{HN5}
\end{align}
where $\theta_m = m \pi / (L+1)$ with $m \in [1,L]$, as $m=L+1$ corresponds to the trivial solution. 
Therefore we have $|\beta_1| = |\beta_2| = r$. 
Choosing $\beta_1 = r e^{i \theta^{(1)}}, \beta_2 = r e^{i \theta^{(2)}}$ and solving $\theta^{(1)} + \theta^{(2)} =  2 \theta_m$ and $\theta^{(1)} - \theta^{(2)} =  2 \theta$ gives us the solution presented in equation (\ref{eq:alpha})
\begin{align}
    \ket{\psi_m} = \mathcal{N}_m \sum_j \lambda^{\frac{j}{2}} \sin{\frac{m \pi}{L+1}} \ket{j} = \mathcal{N}_m \sum_j r^j e^{ij\theta} \ \frac{\omega_m - \omega_m^{-1}}{2 i} \ket{j}, 
\end{align}
with $\omega_m = e^{\frac{i m \pi}{L+1}}$, and
\begin{align}
    \mathcal{N}_m = \frac{2}{\sqrt{ 2 \frac{|\lambda|^2 (1 {-} |\lambda|^L)}{1 {-} |\lambda|} {-} \frac{(|\lambda| e^{i \frac{2\pi m}{L+1}})^2 (1 {-} |\lambda|^L e^{i \frac{2\pi m L}{L+1}})}{1 {-} |\lambda|e^{i \frac{2\pi m}{L+1}}} {-} \frac{(|\lambda| e^{-i \frac{2\pi m}{L+1}})^2 (1 {-} |\lambda|^L e^{-i \frac{2\pi m L}{L+1}})}{1 {-} |\lambda| e^{-i \frac{2\pi m}{L+1}}} }} 
    = \frac{2}{\sqrt{ 2 \frac{r^2(1 {-} r^{2L})}{1 {-} r^{2}} {-} \frac{(r \omega_m)^{2}(1 {-} (r \omega_m)^{2L})}{1 {-} (r \omega_m)^{2}} {-} \frac{(r / \omega_m)^{2}(1 {-} (r / \omega_m)^{2L})}{1 {-} (r / \omega_m)^{2}} }} .
\end{align}
Recalling the definition of QFI
\begin{align}
    F_Q (\lambda) = 4\left(\braket{\partial_\lambda \psi_m|\partial_\lambda \psi_m} - |\braket{\partial_\lambda \psi_m|\psi_m}|^2 \right),
\end{align}
we first calculate $\ket{\partial_\lambda \psi_m}$. Employing the chain rule of partial differentiation with the variables ($r, \theta$), we write down
\begin{align}
    \ket{\partial_\lambda \psi_m} = (\partial_\lambda r) \ket{\partial_r \psi_m} + (\partial_\lambda \theta) \ket{\partial_{\theta} \psi_m} .
    \label{HN_diff1}
\end{align}
Now we express $\ket{\psi_m} = \ket{\tilde{\psi}_m} / || \ket{\tilde{\psi}_m} ||$, with $\ket{\tilde{\psi}_m} = \sum_j (r e^{i \theta})^j \sin{(m \pi/(L+1))}\ket{j}$, and observe that $\ket{\partial_{\theta} \tilde{\psi}_m} = i r \ket{\partial_{r} \tilde{\psi}_m}$. As $|| \ket{\tilde{\psi}_m} ||$ is independent of $\theta$, we have 
\begin{align}
    \ket{\partial_\theta \psi_m} = \frac{1}{|| \ket{\tilde{\psi}_m} ||} \ket{\partial_{\theta} \tilde{\psi}_m} 
    = \frac{i r}{|| \ket{\tilde{\psi}_m} ||} \ket{\partial_{r} \tilde{\psi}_m}
    = \frac{i r}{|| \ket{\tilde{\psi}_m} ||} \partial_{r} (|| \ket{\tilde{\psi}_m} || \ \ket{\psi_m})
    = i r \Big( \partial_r \ket{\psi_m} + \frac{\partial_r || \ket{\tilde{\psi}_m} ||}{|| \ket{\tilde{\psi}_m} ||} \ket{\psi_m} \Big) .
    \label{HN_diff2}
\end{align}
Inserting equation \eqref{HN_diff2} in equation \eqref{HN_diff1}, we get
\begin{align}
    \ket{\partial_\lambda \psi_m} &= (\partial_\lambda r + i r \partial_\lambda \theta) \ket{\partial_r \psi_m} + \Big(i r \frac{\partial_r || \ket{\tilde{\psi}_m} ||}{|| \ket{\tilde{\psi}_m} ||} \partial_\lambda \theta \Big) \ket{\psi_m} \nonumber \\
    &\equiv a \ket{\partial_r \psi_m} + b \ket{\psi_m}.
    \label{HN_diff3}
\end{align}
Using equation \eqref{HN_diff2}, the QFI can be rewritten as
\begin{align}
    F_Q (\lambda) &= 4 |a|^2 \left(\braket{\partial_r \psi_m|\partial_\lambda \psi_m} - |\braket{\partial_r \psi_m|\psi_m}|^2 \right) \nonumber \\
    &= 4 \Big((\partial_\lambda r)^2 + r^2(\partial_\lambda \theta)^2 \Big) \left(\braket{\partial_r \psi_m|\partial_r \psi_m} - |\braket{\partial_r \psi_m|\psi_m}|^2 \right) .
\end{align}
The pre-factor $((\partial_\lambda r)^2 + r^2(\partial_\lambda \theta)^2 )$ approaches a constant value as $\lambda \to \lambda_c$, and does not depend on the system size. The $L$-dependence comes from derivatives with respect to $r$. We can write  
\begin{align}
    F_Q (\lambda) &= \Big((\partial_\lambda r)^2 + r^2(\partial_\lambda \theta)^2 \Big) \nonumber \\
    & \Bigg( \Bigg( 2 (\partial_r \mathcal{N}_m)^2 \sum_{j=1}^L \Big((r)^{2j} {-} (r \omega)^{2j} {-} (r/\omega)^{2j}\Big) \nonumber \\
    & + \frac{\mathcal{N}_m^2}{r^2} \sum_{j=1}^L j^2 \Big((r)^{2j} {-} (r \omega)^{2j} {-} (r/\omega)^{2j}\Big) \nonumber \\
    & + 4 \frac{\mathcal{N}_m \ \partial_r \mathcal{N}_m}{r} \sum_{j=1}^L j \Big((r)^{2j} {-} (r \omega)^{2j} {-} (r/\omega)^{2j}\Big) \Bigg) \nonumber \\
    & - \Bigg| 2 (\mathcal{N}_m \ \partial_r \mathcal{N}_m) \sum_{j=1}^L \Big((r)^{2j} {-} (r \omega)^{2j} {-} (r/\omega)^{2j}\Big)
      + 2 \frac{\mathcal{N}_m^2}{r} \sum_{j=1}^L j \Big((r)^{2j} {-} (r \omega)^{2j} {-} (r/\omega)^{2j}\Big) \Bigg|^2 \Bigg) .
    \label{HN6}
\end{align}
As the transition, $\lambda {\to} \lambda_c$, and consequently, $r {\to} 1$. The $L$-dependence can be shown in the limiting values: $\mathcal{N}_m {\to} \sqrt{2/L} $, $\ \partial_r \mathcal{N}_m {\to} -\sqrt{L/2}$, $\ \sum_j (r)^{2j} {\to} L$, $\ \sum_j (r\omega_m)^{2j} {\to} -1$, $\ \sum_j (r/\omega_m)^{2j} {\to} -1$, $\ \sum_j j(r)^{2j} {\to} L(L+1)/2$, $\ \sum_j j(r\omega_m)^{2j} {\to} (L+1)/(\omega_m^2 - 1)$, $\ \sum_j j(r/\omega_m)^{2j} {\to} (L+1)/(\omega_m^{-2} - 1)$, $\ \sum_j j^2(r)^{2j} {\to} (2L^3 + 3L^2 + L)/6$, $\ \sum_j j^2 (r\omega_m)^{2j} {\to} (L^2)/(\omega_m^2 - 1)$, $\ \sum_j j^2 (r/\omega_m)^{2j} {\to} L^2/(\omega_m^{-2} - 1)$. 
Combining all these terms, the QFI shows $L^2$ scaling in large $L$ limit.

\section{Point gap closing for non-Hermitian SSH model}
\label{Appendix_SSH}

The NH extension of the SSH model is considered with asymmetric intra-cell hopping.
The model still has sublattice symmetry.
The Hamiltonian is given by the equation \eqref{SSH_SL_ham}. 
The model is schematically depicted in figure~\ref{Schematic}(b).
The ansatz for a bulk eigenstate within the GBZ formalism in the presence of NH skin effect for OBC is 
\begin{align}
    \ket{\psi} = \sum_{j} \beta^j (\phi_A \ket{j,A} + \phi_B \ket{j,B}).
    \label{SSH1}
\end{align}
The eigen-equation $H_{\rm SSH} \ket{\psi} = E \ket{\psi}$ in the bulk gives
\begin{align}
    J_2 J_{1L} \beta^2 + (J_2^2 + J_{1L} J_{1R} -E^2) \beta + J_2 J_{1R} = 0 .
    \label{SSH2}
\end{align}
The two solutions $\beta_1$ and $\beta_2$ satisfy
\begin{align}
    \beta_1 \beta_2 = \frac{J_{1R}}{J_{1L}} = r^2 e^{i 2 \theta} \quad \text{and} \quad \beta_1 + \beta_2 = \frac{E^2 - J_2^2 - J_{1L} J_{1R}}{J_2 J_{1L}} .
    \label{SSH3}
\end{align}
The full solution is
\begin{align}
    \ket{\psi} &= \sum_{j} \Big( (c_1 \beta_1^j \phi_{1,A} {+} c_2 \beta_2^j \phi_{2,A}) \ket{j,A} {+} (c_1 \beta_1^j \phi_{1,B} {+} c_2 \beta_2^j \phi_{2,B})  \ket{j,B} \Big) \nonumber \\
    & \equiv \sum_{j}  (\psi_{j,A} \ket{j,A} + \psi_{j,B} \ket{j,B}). 
    \label{SSH4}
\end{align}
Using it for the eigen-equation at the boundaries $(j=1,L)$ and using equation \eqref{SSH3} gives $c_1 \phi_{1,B} = - c_2 \phi_{2,B}$ and
\begin{align}
    \left(\frac{\beta_1}{\beta_2}\right)^{L+1} = \frac{J_{1R} + J_2 \beta_1}{J_{1R} + J_2 \beta_2} ,
    \label{SSH5}
\end{align}
with $|\beta_1| = |\beta_2| = r$. 
Near the transition, $J_{1R}/J_{1L} = \pm 1$. However $J_{1R} = J_{1L}$ is the Hermitian case and therefore, we focus on the $J_{1R} = -J_{1L}$ case. 
Here, $\beta_1 \beta_2 = -1$ and we can choose $\beta_1 = r e^{i \theta}, \beta_2 = -r e^{-i \theta}$. 
The solution for $\theta$ is given by the transcendental equation
\begin{eqnarray}
    \cos{(L+1)\theta} + \frac{i J_2}{\sqrt{|J_{1R} J_{1L}|}} \sin{(L\theta)} &=& 0 \quad (\text{for even $L$}) \nonumber \\
    \sin{(L+1)\theta} - \frac{i J_2}{\sqrt{|J_{1R} J_{1L}|}} \cos{(L\theta)} &=& 0 \quad (\text{for odd $L$}) ,
    \label{SSH6}
\end{eqnarray}
and the wavefunction is given by
\begin{eqnarray}
    \psi_{j,A} &\propto& (i r)^j \Big( i\sin{(j(\theta - \pi/2)) + \frac{J_2}{\sqrt{|J_{1R} J_{1L}|}} \cos{(j(\theta - \pi/2)-\theta)}} \Big), \nonumber \\
    \psi_{j,B} &\propto& (i r)^j \Big( \sin{(j(\theta - \pi/2))} \Big).
    \label{SSH7}
\end{eqnarray}
Although we have an analytical expression for the wavefunction, it requires the knowledge of the phase $\theta$ of $\beta$. 
As this can only be determined by graphically solving the transcendental equations in equation \eqref{SSH6}, the QFI cannot be fully expressed analytically to extract the scaling behavior.
The quadratic scaling of QFI at the transition have been displayed with numerical calculations.
 
\section{Point gap closing for non-Hermitian 2D Chern insulator}
\label{Appendix_2D}

\begin{figure*}[t]
\centering
\includegraphics[width = 0.99\linewidth]{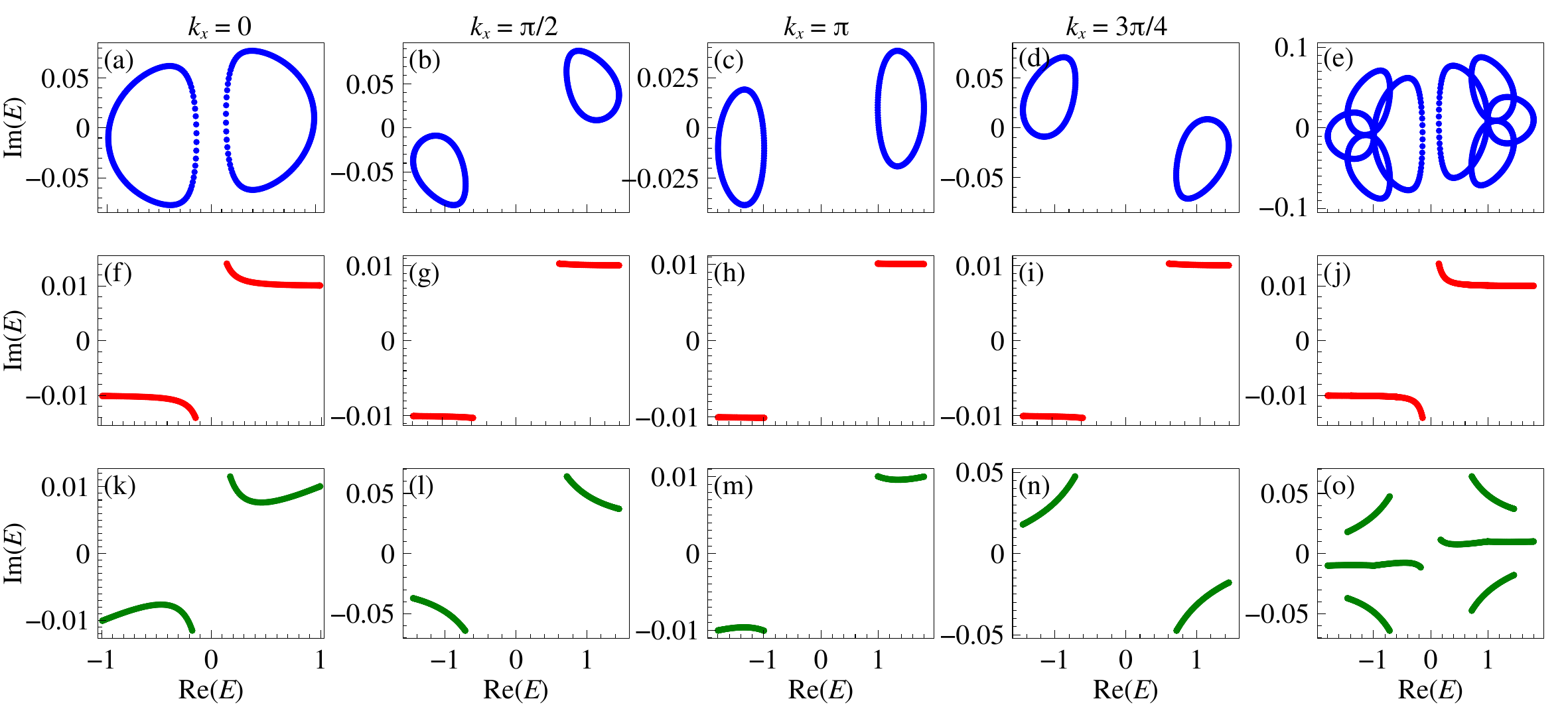}
\caption{Decomposition of PBC spectrum of the 2D Chern insulator model. Few $k_x$ points are chosen to show the energy along the Brillouin zone of $k_y$, forming spectral loops or arcs. The fixed parameters for a $200 \times 200$ lattice are, $m_z = 1, t_2 = 0.2, \gamma_x = 0.1, \gamma_z = 0.01$. (top panel) $t_1 = 0.2, \gamma_y = 0.1$. (middle panel) $t_1 = 0, \gamma_y = 0.1$. (bottom panel) $t_1 = 0.2, \gamma_y = 0$.}
\label{Fig_2D}
\end{figure*}

A NH extension of a prototypical 2D Chern insulator is considered with imaginary Zeeman terms.
The Bloch Hamiltonian for a square lattice with internal levels $a$ and $b$ is given by the equation \eqref{Ch_ham}. 
In real space the model is schematically depicted in figure~\ref{Schematic}(c).
For this $L {\times} L$ lattice, the PBC spectral area can be decomposed into $L$ number of loops in the complex energy plane, each one corresponding to a particular $k_x$ value. 
This is shown for some representative $k_x$ points in the top panel of figure~\ref{Fig_2D} where figure~\ref{Fig_2D}(e) shows how they add up to generate the total spectral area. 
With suitable choice of Hamiltonian parameters, all the loops can be contracted to to arcs (middle panel of figure~\ref{Fig_2D}) in such a way that the overall spectrum also has zero area, as shown in figure~\ref{Fig_2D}(j). 
This causes the NH skin effect to vanish irrespective of OBC geometry. 
The bottom panel shows another type of shrinkage to arcs which do not add up to an overall arc for the full system and the resulting spectrum still has finite area (figure~\ref{Fig_2D}(o)). 
In this case, NH skin effect vanishes when OBC is taken only along the $y$-direction.
To systematically locate the critical parameters for which the two different types of NH Skin effect vanishing occur, we look at the point gap topology of a spectral loop formed at a fixed value of $k_x$.
In this context, the GBZ ansatz for a bulk eigenstate within the GBZ formalism in the presence of NH skin effect for OBC only along $y$-direction is
\begin{align}
    \ket{\psi} = \sum_{j} \beta^j (\phi_a \ket{j,a} + \phi_b \ket{j,b}),
    \label{CH1}
\end{align}
where $a, b$ are the two internal states. 
The eigen-equation for the $k_x$-dependent quasi-1D Hamiltonian gives
\begin{align}
    &  (t_1^2 - t_2^2) \beta^4 - 2 (\gamma_y t_1 + (m_z {-} 2t_2\cos{k_x} {+} i\gamma_z) t_2) \beta^3 \nonumber \\
    +& (E^2 {+} \gamma_y^2 + 2 (t_1^2 {-} t_2^2) {-} (m_z {-} 2t_2\cos{k_x} {+} i\gamma_z)^2 {-} (2 t_1 \sin{k_x} {+} i \gamma_x)^2) \beta^2 \nonumber \\
    +& 2 (\gamma_y t_1 - (m_z {-} 2t_2\cos{k_x} {+} i\gamma_z) t_2) \beta + (t_1^2 - t_2^2) = 0 .
    \label{CH2}
\end{align}
This is a quartic equation, but for the special case of $t_1 {=} t_2$, reduces to a quadratic one with the two solutions $\beta_1$ and $\beta_2$ satisfying
\begin{align}
    \beta_1 \beta_2 = \frac{m_z -2t_2\cos{k_x} +i\gamma_z - \gamma_y}{m_z -2t_2\cos{k_x} +i\gamma_z + \gamma_y} .
    \label{CH3}
\end{align}
With the constraint $|\beta_1| = |\beta_2|$, it is easy to see that for $\gamma_y = 0$, the NH skin effect vanishes for the stripe geometry. 
This observation is in fact a known result~\cite{Yao2018Non} and correspond to the bottom panel of figure~\ref{Fig_2D}.
However, we numerically confirm that this also holds for $t_1 \ne t_2$.
The effective 1D Hamiltonian along $y$-direction resembles a SSH chain with the asymmetric inter-cell hoppings $\tilde{J}_{1L} = m_z - 2 t_2 \cos{k_x} + i \gamma_z + \gamma_y$, $\tilde{J}_{1R} = m_z - 2 t_2 \cos{k_x} + i \gamma_z - \gamma_y$, intra-cell hopping $\tilde{J}_2 = -(t_1 + t_2)$, hopping between site $a$ and site $b$ in the right-adjacent cell $\tilde{J}_3 = t_1 - t_2$, and on-site gain and loss terms with strength $2 t_1 \sin{k_x} + i \gamma_x$.
For this type of SSH chain, the point gap closing occurs at $\tilde{J}_2 = \pm \tilde{J}_3$, which gives $t_1 = 0$ and $t_2 = 0$.
We numerically observe that the first case correspond to the total spectral area going to zero (middle panel of figure~\ref{Fig_2D}) where as the second case is similar to that shown in the bottom panel of figure~\ref{Fig_2D}.

\section{Classical Fisher information in position basis}
\label{Appendix_CFI}

\begin{figure*}[t]
\centering
\includegraphics[width = 0.99\linewidth]{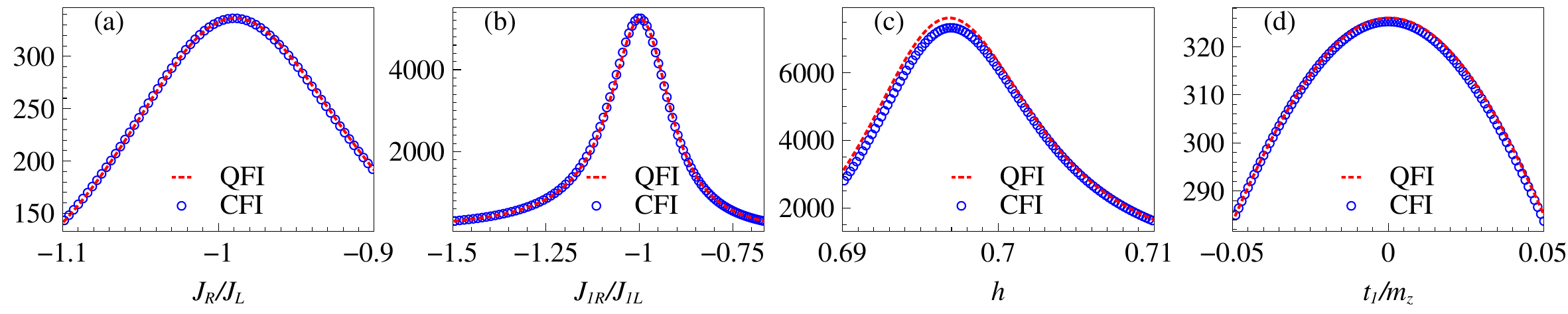}
\caption{Comparison of QFI and CFI for the different point-gapped models considered in this work. (a) Hatano-Nelson model on a 100 site chain. (b) NH version of SSH model on a 100 site chain with $J_2 {= }1$. (c) NH extension of AAH model on a 610 site chain with $J {=} V {=} 1$, $\alpha {=} (\sqrt{5}{+}1)/2$, $\theta {=} 0$. (d) 2D Chern insulator  on a $70 \times 70$ lattice with $m_z {=} 1, t_2 {=} 0.2, \gamma_x {=} 0.1, \gamma_z {=} 0.01$.}
\label{Fig_CFI}
\end{figure*}

The calculation of the classical Fisher information (CFI) in position basis follows the procedure mentioned in section~\ref{sensing}.
The CFI and QFI are quite close to each other for the Hatano-Nelson model, the NH version of SSH model, and the 2D Chern insulator model, as shown in figures~\ref{Fig_CFI} (a), (b), and (d), respectively. 
For the NH extension of AAH model, the discrepancy between CFI and QFI near the transition (figure~\ref{Fig_CFI} (c)) goes down as one goes away from the transition, although the scaling behaviour for CFI and QFI reamin unchanged.

\section*{References}
\bibliographystyle{iopart-num}  
\bibliography{Paper}

\providecommand{\newblock}{}
\begin{thebibliography}{10}
\expandafter\ifx\csname url\endcsname\relax
  \def\url#1{{\tt #1}}\fi
\expandafter\ifx\csname urlprefix\endcsname\relax\def\urlprefix{URL }\fi
\providecommand{\eprint}[2][]{\url{#2}}

\bibitem{Ashida2020Non}
Ashida Y, Gong Z and Ueda M 2020 {\em Advances in Physics\/}
  \href{http://dx.doi.org/10.1080/00018732.2021.1876991}{{\bf 69} 249--435}

\bibitem{Bergholtz2021Exceptional}
Bergholtz E~J, Budich J~C and Kunst F~K 2021 {\em Rev. Mod. Phys.\/}
  \href{http://dx.doi.org/10.1103/RevModPhys.93.015005}{{\bf 93}(1) 015005}

\bibitem{Okuma2023Non}
Okuma N and Sato M 2023 {\em Annual Review of Condensed Matter Physics\/}
  \href{http://dx.doi.org/10.1146/annurev-conmatphys-040521-033133}{{\bf 14}
  83--107}

\bibitem{Kawabata2019Symmetry}
Kawabata K, Shiozaki K, Ueda M and Sato M 2019 {\em Phys. Rev. X\/}
  \href{http://dx.doi.org/10.1103/PhysRevX.9.041015}{{\bf 9}(4) 041015}

\bibitem{Wiersig2014Enhancing}
Wiersig J 2014 {\em Phys. Rev. Lett.\/}
  \href{http://dx.doi.org/10.1103/PhysRevLett.112.203901}{{\bf 112}(20) 203901}

\bibitem{Liu2016Metrology}
Liu Z~P, Zhang J, \"Ozdemir i~m~c~K, Peng B, Jing H, L\"u X~Y, Li C~W, Yang L,
  Nori F and Liu Y~x 2016 {\em Phys. Rev. Lett.\/}
  \href{http://dx.doi.org/10.1103/PhysRevLett.117.110802}{{\bf 117}(11) 110802}

\bibitem{Langbein2018No}
Langbein W 2018 {\em Phys. Rev. A\/}
  \href{http://dx.doi.org/10.1103/PhysRevA.98.023805}{{\bf 98}(2) 023805}

\bibitem{Lau2018Fundamental}
Lau H~K and Clerk A~A 2018 {\em Nature Communications\/}
  \href{http://dx.doi.org/10.1038/s41467-018-06477-7}{{\bf 9} 4320} ISSN
  2041-1723

\bibitem{Zhang2019Quantum}
Zhang M, Sweeney W, Hsu C~W, Yang L, Stone A~D and Jiang L 2019 {\em Phys. Rev.
  Lett.\/} \href{http://dx.doi.org/10.1103/PhysRevLett.123.180501}{{\bf
  123}(18) 180501}

\bibitem{Chen2019Sensitivity}
Chen C, Jin L and Liu R~B 2019 {\em New Journal of Physics\/}
  \href{http://dx.doi.org/10.1088/1367-2630/ab32ab}{{\bf 21} 083002}

\bibitem{budich2020non}
Budich J~C and Bergholtz E~J 2020 {\em Phys. Rev. Lett.\/} {\bf 125} 180403

\bibitem{Koch2022Quantum}
Koch F and Budich J~C 2022 {\em Phys. Rev. Research\/}
  \href{http://dx.doi.org/10.1103/PhysRevResearch.4.013113}{{\bf 4}(1) 013113}

\bibitem{Schomerus2020Nonreciprocal}
Schomerus H 2020 {\em Phys. Rev. Research\/}
  \href{http://dx.doi.org/10.1103/PhysRevResearch.2.013058}{{\bf 2}(1) 013058}

\bibitem{McDonald2020Exponentially}
McDonald A and Clerk A~A 2020 {\em Nature Communications\/}
  \href{http://dx.doi.org/10.1038/s41467-020-19090-4}{{\bf 11} 5382} ISSN
  2041-1723

\bibitem{Hodaei2017Enhanced}
Hodaei H, Hassan A~U, Wittek S, Garcia-Gracia H, El-Ganainy R, Christodoulides
  D~N and Khajavikhan M 2017 {\em Nature\/}
  \href{http://dx.doi.org/10.1038/nature23280}{{\bf 548} 187--191} ISSN
  1476-4687

\bibitem{Yu2020Experimental}
Yu S, Meng Y, Tang J~S, Xu X~Y, Wang Y~T, Yin P, Ke Z~J, Liu W, Li Z~P, Yang
  Y~Z, Chen G, Han Y~J, Li C~F and Guo G~C 2020 {\em Phys. Rev. Lett.\/}
  \href{http://dx.doi.org/10.1103/PhysRevLett.125.240506}{{\bf 125}(24) 240506}

\bibitem{Wang2020Petermann}
Wang H, Lai Y~H, Yuan Z, Suh M~G and Vahala K 2020 {\em Nature
  Communications\/} \href{http://dx.doi.org/10.1038/s41467-020-15341-6}{{\bf
  11} 1610} ISSN 2041-1723

\bibitem{Edvardsson2022Sensitivity}
Edvardsson E and Ardonne E 2022 {\em Phys. Rev. B\/}
  \href{http://dx.doi.org/10.1103/PhysRevB.106.115107}{{\bf 106}(11) 115107}

\bibitem{Ding2023Fundamental}
Ding W, Wang X and Chen S 2023 {\em Phys. Rev. Lett.\/}
  \href{http://dx.doi.org/10.1103/PhysRevLett.131.160801}{{\bf 131}(16) 160801}

\bibitem{paris2009quantum}
Paris M~G 2009 {\em Int. J. Quantum Inf.\/} {\bf 07} 125--137

\bibitem{giovannetti2004quantum}
Giovannetti V, Lloyd S and Maccone L 2004 {\em Science\/} {\bf 306} 1330--1336

\bibitem{giovannetti2006quantum}
Giovannetti V, Lloyd S and Maccone L 2006 {\em Phys. Rev. Lett.\/} {\bf 96}
  010401

\bibitem{frowis2011stable}
Fr{\"o}wis F and D{\"u}r W 2011 {\em Phys. Rev. Lett.\/} {\bf 106} 110402

\bibitem{PhysRevA.97.042112}
Wang K, Wang X, Zhan X, Bian Z, Li J, Sanders B~C and Xue P 2018 {\em Phys.
  Rev. A\/} \href{http://dx.doi.org/10.1103/PhysRevA.97.042112}{{\bf 97}(4)
  042112}

\bibitem{kwon2019nonclassicality}
Kwon H, Tan K~C, Volkoff T and Jeong H 2019 {\em Phys. Rev. Lett.\/} {\bf 122}
  040503

\bibitem{demkowicz2012elusive}
Demkowicz-Dobrza{\'n}ski R, Ko{\l}ody{\'n}ski J and Gu{\c{t}}{\u{a}} M 2012
  {\em Nat. Commun.\/} {\bf 3} 1063

\bibitem{Pasquale2013Quantum}
De~Pasquale A, Rossini D, Facchi P and Giovannetti V 2013 {\em Phys. Rev. A\/}
  \href{http://dx.doi.org/10.1103/PhysRevA.88.052117}{{\bf 88}(5) 052117}

\bibitem{zanardi2006ground}
Zanardi P and Paunkovi{\'c} N 2006 {\em Phys. Rev. E\/} {\bf 74} 031123

\bibitem{zanardi2007mixed}
Zanardi P, Quan H, Wang X and Sun C 2007 {\em Phys. Rev. A\/} {\bf 75} 032109

\bibitem{zanardi2008quantum}
Zanardi P, Paris M~G and Venuti L~C 2008 {\em Phys. Rev. A\/} {\bf 78} 042105

\bibitem{invernizzi2008optimal}
Invernizzi C, Korbman M, Venuti L~C and Paris M~G 2008 {\em Phys. Rev. A\/}
  {\bf 78} 042106

\bibitem{gu2010fidelity}
Gu S~J 2010 {\em Int. J. Mod. Phys. B\/} {\bf 24} 4371--4458

\bibitem{gammelmark2011phase}
Gammelmark S and M{\o}lmer K 2011 {\em New J. Phys.\/} {\bf 13} 053035

\bibitem{skotiniotis2015quantum}
Skotiniotis M, Sekatski P and D{\"u}r W 2015 {\em New J. Phys.\/} {\bf 17}
  073032

\bibitem{rams2018limits}
Rams M~M, Sierant P, Dutta O, Horodecki P and Zakrzewski J 2018 {\em Phys. Rev.
  X\/} {\bf 8} 021022

\bibitem{chu2021dynamic}
Chu Y, Zhang S, Yu B and Cai J 2021 {\em Phys. Rev. Lett.\/} {\bf 126} 010502

\bibitem{liu2021experimental}
Liu R, Chen Y, Jiang M, Yang X, Wu Z, Li Y, Yuan H, Peng X and Du J 2021 {\em
  npj Quantum Inf.\/} {\bf 7} 1--7

\bibitem{montenegro2021global}
Montenegro V, Mishra U and Bayat A 2021 {\em Phys. Rev. Lett.\/} {\bf 126}
  200501

\bibitem{mishra2021driving}
Mishra U and Bayat A 2021 {\em Phys. Rev. Lett.\/} {\bf 127} 080504

\bibitem{Mishra2022Integrable}
Mishra U and Bayat A 2022 {\em Scientific Reports\/}
  \href{http://dx.doi.org/10.1038/s41598-022-17381-y}{{\bf 12} 14760} ISSN
  2045-2322

\bibitem{he2023stark}
He X, Yousefjani R and Bayat A 2023 {\em Phys. Rev. Lett.\/}
  \href{http://dx.doi.org/10.1103/PhysRevLett.131.010801}{{\bf 131}(1) 010801}

\bibitem{yousefjani2023long}
Yousefjani R, He X and Bayat A 2023 {\em Chinese Physics B\/} {\bf 32} 100313

\bibitem{montenegro2023quantumenhanced}
Montenegro V, Genoni M~G, Bayat A and Paris M~G~A 2023 {\em Communications
  Physics\/} \href{http://dx.doi.org/10.1038/s42005-023-01423-6}{{\bf 6} 304}
  ISSN 2399-3650

\bibitem{iemini2023floquet}
Iemini F, Fazio R and Sanpera A 2023 {\em arXiv:2306.03927\/}

\bibitem{Sarkar2022Free}
Sarkar S, Mukhopadhyay C, Alase A and Bayat A 2022 {\em Phys. Rev. Lett.\/}
  \href{http://dx.doi.org/10.1103/PhysRevLett.129.090503}{{\bf 129}(9) 090503}

\bibitem{Alipor2014Quantum}
Alipour S, Mehboudi M and Rezakhani A~T 2014 {\em Phys. Rev. Lett.\/}
  \href{http://dx.doi.org/10.1103/PhysRevLett.112.120405}{{\bf 112}(12) 120405}

\bibitem{Yu2023Quantum}
Yu X and Zhang C 2023 {\em Phys. Rev. A\/}
  \href{http://dx.doi.org/10.1103/PhysRevA.108.022215}{{\bf 108}(2) 022215}

\bibitem{Xiao2020Non}
Xiao L, Deng T, Wang K, Zhu G, Wang Z, Yi W and Xue P 2020 {\em Nature
  Physics\/} \href{http://dx.doi.org/10.1038/s41567-020-0836-6}{{\bf 16}
  761--766} ISSN 1745-2481

\bibitem{Yu2024Toward}
Yu X, Zhao X, Li L, Hu X~M, Duan X, Yuan H and Zhang C 2024 {\em Science
  Advances\/} \href{http://dx.doi.org/10.1126/sciadv.adk7616}{{\bf 10}
  eadk7616}

\bibitem{braunstein1994statistical}
Braunstein S~L and Caves C~M 1994 {\em Phys. Rev. Lett.\/} {\bf 72} 3439

\bibitem{Hasan2010Colloquium}
Hasan M~Z and Kane C~L 2010 {\em Rev. Mod. Phys.\/}
  \href{http://dx.doi.org/10.1103/RevModPhys.82.3045}{{\bf 82}(4) 3045--3067}

\bibitem{Yao2018Edge}
Yao S and Wang Z 2018 {\em Phys. Rev. Lett.\/}
  \href{http://dx.doi.org/10.1103/PhysRevLett.121.086803}{{\bf 121}(8) 086803}

\bibitem{Martinez2018Non}
Martinez~Alvarez V~M, Barrios~Vargas J~E and Foa~Torres L~E~F 2018 {\em Phys.
  Rev. B\/} \href{http://dx.doi.org/10.1103/PhysRevB.97.121401}{{\bf 97}(12)
  121401}

\bibitem{Borgnia2018Non}
Borgnia D~S, Kruchkov A~J and Slager R~J 2020 {\em Phys. Rev. Lett.\/}
  \href{http://dx.doi.org/10.1103/PhysRevLett.124.056802}{{\bf 124}(5) 056802}

\bibitem{Okuma2020Topological}
Okuma N, Kawabata K, Shiozaki K and Sato M 2020 {\em Phys. Rev. Lett.\/}
  \href{http://dx.doi.org/10.1103/PhysRevLett.124.086801}{{\bf 124}(8) 086801}

\bibitem{Zhang2020Correspondence}
Zhang K, Yang Z and Fang C 2020 {\em Phys. Rev. Lett.\/}
  \href{http://dx.doi.org/10.1103/PhysRevLett.125.126402}{{\bf 125}(12) 126402}

\bibitem{Zhang2022Universal}
Zhang K, Yang Z and Fang C 2022 {\em Nature Communications\/}
  \href{http://dx.doi.org/10.1038/s41467-022-30161-6}{{\bf 13} 2496} ISSN
  2041-1723

\bibitem{zhong2021nontrivial}
Zhong J, Wang K, Park Y, Asadchy V, Wojcik C~C, Dutt A and Fan S 2021 {\em
  Phys. Rev. B\/} \href{http://dx.doi.org/10.1103/PhysRevB.104.125416}{{\bf
  104}(12) 125416}

\bibitem{Gong2018Topological}
Gong Z, Ashida Y, Kawabata K, Takasan K, Higashikawa S and Ueda M 2018 {\em
  Phys. Rev. X\/} \href{http://dx.doi.org/10.1103/PhysRevX.8.031079}{{\bf 8}(3)
  031079}

\bibitem{Lieu2019Non}
Lieu S 2019 {\em Phys. Rev. B\/}
  \href{http://dx.doi.org/10.1103/PhysRevB.100.085110}{{\bf 100}(8) 085110}

\bibitem{Panda2020Entanglement}
Panda A and Banerjee S 2020 {\em Phys. Rev. B\/}
  \href{http://dx.doi.org/10.1103/PhysRevB.101.184201}{{\bf 101}(18) 184201}

\bibitem{Banerjee2022Chiral}
Banerjee A, Hegde S~S, Agarwala A and Narayan A 2022 {\em Phys. Rev. B\/}
  \href{http://dx.doi.org/10.1103/PhysRevB.105.205403}{{\bf 105}(20) 205403}

\bibitem{Kawabata2020Higher}
Kawabata K, Sato M and Shiozaki K 2020 {\em Phys. Rev. B\/}
  \href{http://dx.doi.org/10.1103/PhysRevB.102.205118}{{\bf 102}(20) 205118}

\bibitem{Hatano1996Localization}
Hatano N and Nelson D~R 1996 {\em Phys. Rev. Lett.\/}
  \href{http://dx.doi.org/10.1103/PhysRevLett.77.570}{{\bf 77}(3) 570--573}

\bibitem{Hatano1997Vortex}
Hatano N and Nelson D~R 1997 {\em Phys. Rev. B\/}
  \href{http://dx.doi.org/10.1103/PhysRevB.56.8651}{{\bf 56}(14) 8651--8673}

\bibitem{Yao2018Non}
Yao S, Song F and Wang Z 2018 {\em Phys. Rev. Lett.\/}
  \href{http://dx.doi.org/10.1103/PhysRevLett.121.136802}{{\bf 121}(13) 136802}

\bibitem{Yokomizo2019Non}
Yokomizo K and Murakami S 2019 {\em Phys. Rev. Lett.\/}
  \href{http://dx.doi.org/10.1103/PhysRevLett.123.066404}{{\bf 123}(6) 066404}

\bibitem{Yokomizo2023Non}
Yokomizo K and Murakami S 2023 {\em Phys. Rev. B\/}
  \href{http://dx.doi.org/10.1103/PhysRevB.107.195112}{{\bf 107}(19) 195112}

\bibitem{guo2021exact}
Guo C~X, Liu C~H, Zhao X~M, Liu Y and Chen S 2021 {\em Phys. Rev. Lett.\/}
  \href{http://dx.doi.org/10.1103/PhysRevLett.127.116801}{{\bf 127}(11) 116801}

\bibitem{Halder2023Properties}
Halder D, Ganguly S and Basu S 2022 {\em Journal of Physics: Condensed
  Matter\/} \href{http://dx.doi.org/10.1088/1361-648X/acadc5}{{\bf 35} 105901}

\bibitem{Longhi2019Topological}
Longhi S 2019 {\em Phys. Rev. Lett.\/}
  \href{http://dx.doi.org/10.1103/PhysRevLett.122.237601}{{\bf 122}(23) 237601}

\bibitem{Jiang2019Interplay}
Jiang H, Lang L~J, Yang C, Zhu S~L and Chen S 2019 {\em Phys. Rev. B\/}
  \href{http://dx.doi.org/10.1103/PhysRevB.100.054301}{{\bf 100}(5) 054301}

\bibitem{wei2019fidelity}
Wei B~B 2019 {\em Phys. Rev. A\/}
  \href{http://dx.doi.org/10.1103/PhysRevA.99.042117}{{\bf 99}(4) 042117}

\bibitem{Shen2018Topological}
Shen H, Zhen B and Fu L 2018 {\em Phys. Rev. Lett.\/}
  \href{http://dx.doi.org/10.1103/PhysRevLett.120.146402}{{\bf 120}(14) 146402}

\bibitem{qi2006topological}
Qi X~L, Wu Y~S and Zhang S~C 2006 {\em Phys. Rev. B\/}
  \href{http://dx.doi.org/10.1103/PhysRevB.74.085308}{{\bf 74}(8) 085308}

\bibitem{Helbig2020Generalized}
Helbig T, Hofmann T, Imhof S, Abdelghany M, Kiessling T, Molenkamp L~W, Lee
  C~H, Szameit A, Greiter M and Thomale R 2020 {\em Nature Physics\/}
  \href{http://dx.doi.org/10.1038/s41567-020-0922-9}{{\bf 16} 747--750} ISSN
  1745-2481

\bibitem{Hofmann2020Reciprocal}
Hofmann T, Helbig T, Schindler F, Salgo N, Brzezi\ifmmode~\acute{n}\else
  \'{n}\fi{}ska M, Greiter M, Kiessling T, Wolf D, Vollhardt A,
  Kaba\ifmmode~\check{s}\else \v{s}\fi{}i A, Lee C~H, Bilu\ifmmode
  \check{s}\else \v{s}\fi{}i\ifmmode~\acute{c}\else \'{c}\fi{} A, Thomale R and
  Neupert T 2020 {\em Phys. Rev. Res.\/}
  \href{http://dx.doi.org/10.1103/PhysRevResearch.2.023265}{{\bf 2}(2) 023265}

\bibitem{Liu2021Non}
Liu S, Shao R, Ma S, Zhang L, You O, Wu H, Xiang Y~J, Cui T~J and Zhang S 2021
  {\em Research\/} \href{http://dx.doi.org/10.34133/2021/5608038}{{\bf 2021}}

\bibitem{Zhang2021Acoustic}
Zhang L, Yang Y, Ge Y, Guan Y~J, Chen Q, Yan Q, Chen F, Xi R, Li Y, Jia D, Yuan
  S~Q, Sun H~X, Chen H and Zhang B 2021 {\em Nature Communications\/}
  \href{http://dx.doi.org/10.1038/s41467-021-26619-8}{{\bf 12} 6297} ISSN
  2041-1723

\bibitem{Gao2022Anomalous}
Gao H, Xue H, Gu Z, Li L, Zhu W, Su Z, Zhu J, Zhang B and Chong Y~D 2022 {\em
  Phys. Rev. B\/} \href{http://dx.doi.org/10.1103/PhysRevB.106.134112}{{\bf
  106}(13) 134112}

\bibitem{Weidemann2020Topological}
Weidemann S, Kremer M, Helbig T, Hofmann T, Stegmaier A, Greiter M, Thomale R
  and Szameit A 2020 {\em Science\/}
  \href{http://dx.doi.org/10.1126/science.aaz8727}{{\bf 368} 311--314}

\bibitem{Song2020Two}
Song Y, Liu W, Zheng L, Zhang Y, Wang B and Lu P 2020 {\em Phys. Rev. Appl.\/}
  \href{http://dx.doi.org/10.1103/PhysRevApplied.14.064076}{{\bf 14}(6) 064076}

\bibitem{Brandenbourger2019Non}
Brandenbourger M, Locsin X, Lerner E and Coulais C 2019 {\em Nature
  Communications\/} \href{http://dx.doi.org/10.1038/s41467-019-12599-3}{{\bf
  10} 4608} ISSN 2041-1723

\bibitem{Ghatak2020Observation}
Ghatak A, Brandenbourger M, van Wezel J and Coulais C 2020 {\em Proceedings of
  the National Academy of Sciences\/}
  \href{http://dx.doi.org/10.1073/pnas.2010580117}{{\bf 117} 29561--29568}

\bibitem{Lapp2019Engineering}
Lapp S, Ang’ong’a J, An F~A and Gadway B 2019 {\em New Journal of
  Physics\/} \href{http://dx.doi.org/10.1088/1367-2630/ab1147}{{\bf 21} 045006}

\bibitem{Gou2020Tunable}
Gou W, Chen T, Xie D, Xiao T, Deng T~S, Gadway B, Yi W and Yan B 2020 {\em
  Phys. Rev. Lett.\/}
  \href{http://dx.doi.org/10.1103/PhysRevLett.124.070402}{{\bf 124}(7) 070402}

\bibitem{Xiao2021Observation}
Xiao L, Deng T, Wang K, Wang Z, Yi W and Xue P 2021 {\em Phys. Rev. Lett.\/}
  \href{http://dx.doi.org/10.1103/PhysRevLett.126.230402}{{\bf 126}(23) 230402}

\bibitem{Lin2022Observation}
Lin Q, Li T, Xiao L, Wang K, Yi W and Xue P 2022 {\em Nature Communications\/}
  \href{http://dx.doi.org/10.1038/s41467-022-30938-9}{{\bf 13} 3229} ISSN
  2041-1723

\bibitem{Lin2022ObservationAAH}
Lin Q, Li T, Xiao L, Wang K, Yi W and Xue P 2022 {\em Phys. Rev. Lett.\/}
  \href{http://dx.doi.org/10.1103/PhysRevLett.129.113601}{{\bf 129}(11) 113601}

\bibitem{Mochizuki2016Explicit}
Mochizuki K, Kim D and Obuse H 2016 {\em Phys. Rev. A\/}
  \href{http://dx.doi.org/10.1103/PhysRevA.93.062116}{{\bf 93}(6) 062116}

\bibitem{Lieu2018Topological}
Lieu S 2018 {\em Phys. Rev. B\/}
  \href{http://dx.doi.org/10.1103/PhysRevB.97.045106}{{\bf 97}(4) 045106}

\bibitem{Kunst2018Biorthogonal}
Kunst F~K, Edvardsson E, Budich J~C and Bergholtz E~J 2018 {\em Phys. Rev.
  Lett.\/} \href{http://dx.doi.org/10.1103/PhysRevLett.121.026808}{{\bf 121}(2)
  026808}

\end{thebibliography}

\end{document}